\documentclass[pra,twocolumn,amsmath,amssymb,superscriptaddress,floatfix]{revtex4}

\usepackage{graphicx}
\usepackage{epstopdf}
\usepackage{dcolumn}
\usepackage{bm}
\usepackage{color}

\usepackage{epsfig,float,afterpage,amssymb,wrapfig}
\usepackage[caption=false]{subfig}
\usepackage{placeins}
\usepackage[normalem]{ulem}
\usepackage{soul}

\DeclareMathAlphabet{\bi}{OML}{cmm}{b}{it}

\newcommand{\half}{\text{$\textstyle\frac{1}{2}$}}

\begin{document}
\title{Dynamical Kondo effect and Kondo destruction in effective models for quantum-critical heavy fermion metals}

\author{Ang Cai}
\affiliation{Department of Physics and Astronomy, Rice Center for Quantum Materials, Rice University,
Houston, Texas, 77005, USA}
\author{Haoyu Hu}
\affiliation{Department of Physics and Astronomy, Rice Center for Quantum Materials, Rice University,
Houston, Texas, 77005, USA}
\author{Kevin Ingersent}
\affiliation{Department of Physics, University of Florida, Gainesville, Florida, 32611-8440, USA}
\author{Silke Paschen}
\affiliation{Institute of Solid State Physics, Vienna University of Technology, 1040 Vienna, Austria}
\affiliation{Department of Physics and Astronomy, Rice Center for Quantum Materials,  Rice University,
Houston, Texas, 77005, USA}
\author{Qimiao Si}
\affiliation{Department of Physics and Astronomy, Rice Center for Quantum Materials, Rice University,
Houston, Texas, 77005, USA}

\date{\today}

\begin{abstract}
Quantum criticality in certain heavy-fermion metals is believed to go beyond the Landau framework of
order-parameter fluctuations. In particular, there is considerable evidence for Kondo destruction:
a disappearance of the {\it static} Kondo singlet amplitude that results in a sudden
reconstruction of Fermi surface across the quantum critical point and an extra critical energy scale.
This effect can be analyzed in terms of a dynamical interplay between the Kondo and RKKY interactions.
In the Kondo-destroyed phase, a well-defined Kondo resonance is lost, but Kondo singlet
correlations remain at nonzero frequencies. This dynamical effect allows for mass enhancement in the
Kondo-destroyed phase. Here, we elucidate the dynamical Kondo effect in Bose-Fermi Kondo/Anderson
models,  which unambiguously exhibit Kondo-destruction quantum critical points. We show that a simple
physical quantity---the expectation value $\langle {\bf S}_{f} \cdot {\bf s}_{c} \rangle$ for the dot
product of the local ($f$) and conduction-electron ($c$) spins---varies continuously across
such quantum critical points. A nonzero $\langle {\bf S}_{f} \cdot {\bf s}_{c} \rangle$ manifests the
dynamical Kondo effect that operates in the Kondo-destroyed phase. Implications are discussed for the
stability of Kondo-destruction quantum criticality as well as the understanding of experimental results
in quantum critical heavy-fermion metals.
\end{abstract}

\maketitle

\section{Introduction}
\label{sec:introduction}

Quantum criticality is a unifying theme for strongly correlated systems, driving properties such as 
non-Fermi liquid (``strange metal") behavior and unconventional superconductivity
\cite{SpecialIssue2013,SpecialIssue2010,sachdev2011quantum,si2010heavy}.
It often develops in bad metals ({\it i.e.}, metals with strong correlations) 
and near antiferromagnetic order.
Antiferromagnetic quantum critical points (QCPs) are especially prevalent in
heavy-fermion metals \cite{si2010heavy,Coleman-Nature,stewart2001non,si2013quantum},
which are highly tunable due to their small energy scales.
One of the overarching questions is whether quantum criticality 
in heavy-fermion systems goes beyond the Landau framework of order-parameter fluctuations,
as proposed in terms of Kondo destruction
\cite{si2001locally,Colemanetal,senthil2004a}.
At a Kondo-destruction QCP, a new type of critical modes arises from the suppression of the
{\it static} Kondo singlet amplitude.
Some of the salient properties associated with Kondo destruction have been experimentally observed,
and are in sharp contrast with the expectations of the conventional spin-density wave QCP
\cite{hertz1976quantum, millis1993effect, moriya2012spin}. These properties include spin dynamics
obeying an anomalous scaling \cite{schroder2000onset}, charge dynamics that are singular with
$\omega/T$ scaling \cite{prochaska2018singular}, as well as a Fermi surface that undergoes a sudden
reconstruction across the QCP
\cite{paschen2004hall,Fri10.2,shishido2005drastic,custers2012destruction,martelli2017sequential}.
In the Kondo-destroyed phase, the Fermi surface is small, its volume being determined by the
number of conduction ($c$) electrons only. In the
Kondo-screened
phase, by contrast, the $c$ electrons near
the Fermi energy are hybridized with $f$ electrons, and there is a large Fermi surface whose volume
is determined by the combined number of $c$ and $f$ electrons.

Microscopically, Kondo destruction in Kondo lattice models 
originates from the dynamical competition between the Kondo and
Ruderman-Kittel-Kasuya-Yosida (RKKY) exchange interactions. We consider the case when both
these interactions are antiferromagnetic.
The Kondo interaction between each local $f$ moment and its on-site conduction electrons lowers
the ground-state energy through the development of Kondo singlet correlations, while the RKKY
interaction does so by establishing singlet correlations between different local moments and the
associated antiferromagnetic ordering.
When the Kondo interaction dominates, the ground state features Kondo entanglement and absence of
long-range antiferromagnetic order, but antiferromagnetic correlations nonetheless remain.
Likewise, when the RKKY interaction prevails, the ground state displays antiferromagnetic order and 
destruction of ground-state Kondo entanglement, but dynamical Kondo correlations remain.
A Kondo-destruction QCP has been realized in Kondo lattice models
when the dynamical competition between the Kondo and RKKY interactions is taken into account
\cite{si2001locally,si2003local,grempel2003locally,
jxzhu2003continuous,glossop2007magnetic,jxzhu2007zero}.
The dynamical Kondo effect helps stabilize a Kondo-destroyed phase and
allows for the quantum phase transition from the Kondo phase to be second order
\cite{si2014kondo,zhu2003continuous}. 

\subsection{Specific motivation}

The dynamical Kondo effect implies that the quasiparticles near the small Fermi surface in the
Kondo-destroyed phase have enhanced mass. This is important for  understanding properties such as the
enhancement of the Sommerfeld coefficient ($C_{\rm el}/T$), cyclotron mass, and quadratic-in-$T$
coefficient $A$ of the electrical resistivity ($\rho = \rho_0 + A T^2$) in the Kondo-destroyed phase
implicated in YbRh$_2$Si$_2$, Au-doped CeCu$_6$, CeRhIn$_5$, and Ce$_3$Pd$_{20}$Si$_6$
\cite{Geg02.1,Loe94.1,shishido2005drastic,custers2012destruction,martelli2017sequential}.

This effect is in contrast to the expectations from a static mean-field picture for the Kondo lattice,
which reduces the Kondo correlations to just a static amplitude, say the condensation amplitude of
a slave boson \cite{Hew97.1}, and ignores their quantum fluctuating part.
Viewed from this static mean-field perspective, destruction of the Kondo effect implies the complete
decoupling of the local-moment and conduction-electron spins, with a vanishing expectation value
$\langle {\bf S}_{f} \cdot {\bf s}_{c} \rangle$ and the absence of mass enhancement for quasiparticles
near the small Fermi surface.

To be more specific, by the dynamical Kondo effect, we refer to antiferromagnetic correlations
at nonzero frequencies between the local-moment and conduction-electron spins in the Kondo-destroyed
phase. Such correlations can be quantified in terms of the cross susceptibility
\begin{equation}
\chi_{Ss}(\tau)=\langle T_{\tau}  
{\bf S}_{f} (\tau) \cdot  {\bf s}_{c} (0)  \rangle .
\end{equation}
The spectral function $\operatorname{Im} \chi_{Ss}(\omega)$ is the imaginary part of the retarded
correlation function $\chi_{Ss}(\omega) = \chi_{Ss}(i \omega_{n} \rightarrow \omega + i 0^{+})$,
with $\chi_{Ss}(i \omega_{n}) = \int_{0}^{\beta} e^{-i \omega_{n} \tau} \chi_{Ss} (\tau) \, d\tau$.
The dynamical Kondo effect means that $\operatorname{Im} \chi_{Ss}(\omega)$ is nonzero for
$\omega \neq 0 $ in the Kondo-destroyed phase.

Consider now $\langle {\bf S}_{f} \cdot  {\bf s}_{c} \rangle $, 
which is the equal-time limit of  $\chi_{Ss}(\tau)$: $\langle {\bf S}_{f} \cdot  {\bf s}_{c}
 \rangle = \chi _{Ss}(\tau \rightarrow 0^{+} )$.
It is related to the spectral function $\operatorname{Im} \chi_{Ss}(\omega)$ through
\begin{equation}
\langle {\bf S}_{f} \cdot  {\bf s}_{c} \rangle 
=
\frac{1 }{\pi}
\int_{-\infty}^{\infty}   n_{B}  (\omega)  \operatorname{Im} \chi_{Ss}(\omega) \, d \omega , 
\label{expectation_value_kk}
\end{equation}
where $n_{B}(\omega)=1/( e^{\beta \omega} -1 )$ is the Bose-Einstein distribution function. 

It follows from Eq.\ \eqref{expectation_value_kk} that the dynamical Kondo effect implies a nonzero
$\langle {\bf S}_{f} \cdot  {\bf s}_{c} \rangle $ 
in the Kondo-destroyed phase.
This motivates us to use 
 $\langle {\bf S}_{f} \cdot  {\bf s}_{c} \rangle $ as a diagnostic for the dynamical Kondo effect.
We therefore calculate this expectation value in well-defined models, for which
the existence of Kondo destruction can be unambiguously established.

An added impetus
 for the present work is the recent development of an SU(2) continuous-time
quantum Monte Carlo (CT-QMC) method \cite{cai2019} (see also Ref.\,\onlinecite{otsuki2013spin}). 
This method is able to reach low-enough temperatures and study Kondo destruction in models with SU(2)
symmetry.

\subsection{Main results}

In this paper, we study the Bose-Fermi Kondo (BFK) and Bose-Fermi Anderson (BFA) models in their
SU(2) spin-symmetric and Ising spin-anisotropic variants.
We provide multiple lines of evidence for the existence of a continuous quantum phase transition
between Kondo-entangled and Kondo-destroyed phases that possess, respectively,
nonzero and zero amplitudes for the static Kondo singlet.
Our results for $\langle {\bf S}_{f} \cdot  {\bf s}_{c} \rangle$ and (in the Ising-anisotropic case)
$\langle S_{f}^z s_{c}^z \rangle$ show that these quantities are nonzero even in the Kondo-destroyed
phase, where the static Kondo effect is absent.
Through Eq. \eqref{expectation_value_kk}, we can conclude that the dynamical Kondo effect operates 
in the Kondo-destroyed phase.

\subsection{Outline of the paper}

Section \ref{sec:models} describes the BFA models with either SU(2) symmetry or Ising anisotropy,
and the corresponding BFK models that arise in the limit of infinite $f$-site
Coulomb repulsion. These models emerge from the
treatment of the Kondo/Anderson lattice models \cite{si2001locally,si2003local,grempel2003locally,
jxzhu2003continuous,glossop2007magnetic,jxzhu2007zero} within the extended dynamical mean-field theory
(EDMFT) \cite{si1996kosterlitz,smith2000spatial,chitra2000effect} and, thus, they represent 
effective models for quantum-critical heavy fermion metals. In the EDMFT,
one focuses on a single lattice site and integrates out all other sites, obtaining
an effective model with a local $f$ level coupled both to a fermionic bath (capturing the
Kondo interaction) and a bosonic bath (representing the dynamical effect of the RKKY interaction). 
The competition between Kondo and RKKY interactions in the lattice model is translated into a
competition between couplings to the fermionic and bosonic baths of the effective models.

There are several ways to systematically examine the quantum phase transitions of the BFA and BFK
models. Analytically, the models can be studied under an $\epsilon$-expansion renormalization-group (RG)
approach.  Here, the system is either in a Kondo-screened or Kondo-destroyed phase, depending on
whether the system flows to a fixed point with strong Kondo coupling (which signifies the development
of nonzero amplitude for the static Kondo singlet in the system's ground state \cite{Hew97.1})
or to a fixed point with vanishing Kondo coupling (which is the unambiguous evidence
for the suppression of the static Kondo singlet amplitude in the ground state). 
Numerically, the models can be studied using CT-QMC and, in cases with Ising anisotropy, also
with the NRG method. These two numerical methods usually complement each other since CT-QMC 
is exact but is limited to temperatures $T>0$, while the NRG can work directly at $T=0$ 
but (for nonzero temperatures) is unreliable in the frequency regime $0 < |\omega| \leq T$.

Section \ref{sec:existence} focuses on the existence of Kondo destruction in the BFA and BFK models. 
The evidence comes from a combination of (i) analytical $\epsilon$-expansion RG analysis of the BFK
models, revealing unstable Kondo-destruction fixed points separating Kondo-screened and Kondo-destroyed
phases; (ii) numerical evidence for the crossing and scaling collapse of the Binder ratio in the
Ising-anisotropic BFA model; (iii) divergence of the fidelity susceptibility in both the Ising-anisotropic
and SU(2)-symmetric BFA; (iv) the bifurcation of the flow of the many-body spectrum in NRG calculations
toward those of the Kondo-screened and Kondo-destroyed fixed points, for the
Ising-anisotropic case; and (v) the vanishing of quasiparticle weight as the Kondo-destruction QCP 
is approached from either phase, as evidenced by the collapse of a crossover energy scale.    

The central results of our work are presented in Sec.\ \ref{sec:sdots}, which demonstrates the
dynamical Kondo effect. We show that  $\langle {\bf S}_f  \cdot {\bf s}_c \rangle$ is nonzero in
the Kondo-destroyed phase, and clarify that this captures the dynamical Kondo effect. In particular, we
demonstrate that this expectation value evolves continuously from the Kondo-screened phase through the
QCP into the Kondo-destroyed phase. Finally, we discuss our results in Sec.\ \ref{sec:discussion} and
summarize the paper in Sec.\ \ref{sec:summary}.

\section{Models and methods}
\label{sec:models}

This section describes the models considered in this work. The SU(2)-symmetric and Ising-anisotropic
BFK models have been studied analytically using the $\epsilon$-expansion RG
\cite{si1996kosterlitz,zhu2002,zarand2002quantum,smith1999non,sengupta2000spin}.
The $\epsilon$-expansion results serve as a guide for our numerical analyses, which are performed
for the BFA model. The BFA model with SU(2) symmetry or Ising anistropy reduces to the corresponding BFK
model in the limit of infinite on-site repulsion between localized ($f$) electrons.

\subsection{Bose-Fermi Kondo models}

The SU(2)-symmetric BFK Hamiltonian can be written
\begin{eqnarray}
{\cal H}_{\text{BFK}}^{\text{SU(2)}}
&=& J \, {\bf S}_{f} \cdot {\bf s}_c 
+ \sum_{p,\sigma} \epsilon_{p} \, c_{p\sigma}^{\dag} \, c_{p\sigma} \\
&+& g \, {\bf S}_{f} \cdot \sum_{p} \left( \vec{\phi}_{p} + \vec{\phi}_{-p}^{\;\dag} \right)
+ \sum_{p} w_{p} \, \vec{\phi}_{p}^{\;\dag} \cdot \vec{\phi}_{p}, \nonumber
\label{H-BFK-SU2}
\end{eqnarray}
where a spin-\half local moment ${\bf S}_{f}$ is coupled both to the on-site spin
${\bf s}_{c}$  of a fermionic bath $c_{p\sigma}$ through Kondo coupling $J$ and to the
displacement of a vector bosonic bath $\vec{\phi}_{p}$ with coupling $g$. Here,
$s_c^{\alpha}=\half \sum_{p,p',\sigma,\sigma'} c^{\dag}_{p\sigma} \,
\tau_{\sigma \sigma'}^{\alpha} \, c_{p'\sigma'}$ for $\alpha=x$, $y$, $z$, with
$\tau_{\sigma \sigma'}^{\alpha}$ being a Pauli matrix.

The Ising-anisotropic BFK model is specified by the Hamiltonian
\begin{eqnarray}
{\cal H}_{\text{BFK}}^{\text{Ising}}
&=& J _{z} \, {S}_{s}^{z} {s}_{c}^{z}
+ J_{\perp} ( {S}_{f}^{x} {s}_c^{x}  + S_{f}^{y} s_{c}^{y})
+ \sum_{p,\sigma} \epsilon_{p} \, c_{p\sigma}^{\dag} \, c_{p\sigma} \nonumber \\
&+&  g \, {S}_{f} ^{z} \sum_{p} \left( {\phi}^{z}_{p} +  {\phi}_{-p}^{z\;\dag} \right)
+ \sum_{p} w_{p} \, {\phi}_{p}^{\dag} \, {\phi}_{p} ,
\label{H-BFK-Ising}
\end{eqnarray}
where the local moment couples to the fermionic bath through longitudinal and perpendicular Kondo couplings
$J_z$ and $J_{\perp}$, and couples via the $z$ component of its spin to a single component bosonic bath
$\phi_p$.
We assume isotropic bare Kondo couplings $J_z = J_{\perp} = J$, but within the $\epsilon$-expansion approach
one must allow for the possible development of anisotropy under RG flow.

\subsection{Bose-Fermi Anderson models}

For the purposes of CT-QMC study, it is preferable to study the BFA counterparts of the models just described.
The SU(2)-symmetric BFA Hamiltonian is 
\begin{eqnarray}
{\cal H}^{\text{SU(2)}}_{\text{BFA}}
&=& \epsilon_{d} \sum_{\sigma} n_{d\sigma} + U n_{d\uparrow} n_{d\downarrow}
+ V \sum_{p,\sigma} \left( d_{\sigma}^{\dag} \, c_{p\sigma} + \text{H.c.} \right) \nonumber  \\
&+& \sum_{p,\sigma} \epsilon_{p} \, c_{p\sigma}^{\dag} \, c_{p\sigma}
+ g \, {\bf S}_{f} \cdot \sum_{p} \left( \vec{\phi}_{p} + \vec{\phi}_{-p}^{\;\dag} \right) \nonumber \\
&+& \sum_{p} w_{p}\, \vec{\phi}_{p}^{\;\dag} \cdot \vec{\phi}_{p}.
\label{H-BFA-SU2}
\end{eqnarray}
with ${\bf S}_{f}=\half \sum_{\sigma,\sigma'} d_{\sigma}^{\dag} \, \bm{\tau}_{\sigma \sigma'}
\, d_{\sigma'}$ and $n_{f\sigma}= d_{\sigma}^{\dag} \, d_{\sigma}$.

The Ising-anisotropic variant of the BFA model has Hamiltonian
\begin{eqnarray}
{\cal H}^{\text{Ising}}_{\text{BFA}}
&=& \epsilon_{d} \sum_{\sigma} n_{d\sigma} + U n_{d\uparrow} n_{d\downarrow}
+ V \sum_{p,\sigma} \left( f_{\sigma}^{\dag} \, c_{p,\sigma} + \text{H.c.} \right) \nonumber \\ 
&+& \sum_{p,\sigma} \epsilon_{p} \, c_{p\sigma}^{\dag} \, c_{p\sigma}
+ g \, {S}_{f}^{z} \sum_{p} \left( {\phi}_{p} + {\phi}_{-p}^{\;\dag} \right) \nonumber \\
&+& \sum_{p} w_{p} \, {\phi}_{p}^{\;\dag} \, {\phi}_{p} .
\label{H-BFA-Ising}
\end{eqnarray}

The SU(2)-symmetric and Ising-anisotropic BFA models reduce to their BFK counterparts in the limit
$U \rightarrow \infty$, $\epsilon_f \rightarrow -\infty$ of suppressed charge fluctuations on the $f$ site.
Unless the limit is taken in such a way that $U=-2\epsilon_d$, the BFK Hamiltonians are supplemented by
a potential scattering term of the form $W \sum_{p,p',\sigma} c_{p\sigma}^{\dag} \, c_{p'\sigma}$.
However, the coupling $W$ proves to be marginal under RG flow, and it can safely be neglected.

\subsection{Choice of fermionic and bosonic baths}

To complete the models specified in Eqs.\ \eqref{H-BFK-SU2}--\eqref{H-BFA-Ising}, it is necessary to
specify the densities of states of the fermionic and bosonic baths.

For the fermionic bath, we set the density of states 
\begin{equation}
\rho_{F}(\epsilon)=\sum_{p} \delta(\epsilon-\epsilon_{p}) = N_0 \Theta (|D-\epsilon|)
\end{equation}
to be constant over a bandwidth $2D$, which leads to a hybridization function
$\Gamma(\epsilon)= \Gamma_{0} \Theta(|D-\epsilon|)$, with $\Gamma_{0}=\pi N_0  V^{2}$.

The density of states for the bosonic bath is chosen to be sub-ohmic with a power-law exponent
$0<s<1$, written as
\begin{equation}
\rho_{B}(\omega)=\sum_{p} \delta(\omega-\omega_{p} )
  = K^{2}_{0} (\omega/\omega_{c})^{s} \Theta(\omega) f(\omega/\omega_{c}),
\end{equation}
with $f(x)$ being an upper cutoff function. In CT-QMC, we choose a soft upper cutoff $f(x)=\exp(-x)$,
and $K_{0}$ is determined from the normalization condition $\int_{0}^{\infty} \rho_{B}(\omega) = 1$.
In the NRG, we choose a hard upper cutoff $f(x)=\Theta(1-x)$.

Throughout the remainder of this paper, we take $D=\omega_c=1$ as our fundamental energy scale.

\section{Kondo-destruction quantum criticality}
\label{sec:existence}

In this section we demonstrate the existence of a Kondo-destruction
quantum phase transition in the BFK and BFA models with SU(2) symmetry
and Ising anisotropy.
 
\subsection{SU(2) symmetry}

We first consider Kondo destruction in Bose-Fermi models that exhibit full SU(2)
spin rotation symmetry.

\subsubsection{$\epsilon$-expansion RG for the BFK model}

To set the stage for our analysis, we briefly recall
the RG analysis via expansion in $\epsilon=1-s$ 
\cite{zhu2002,zarand2002quantum,smith1999non,sengupta2000spin}. 
To order $\epsilon^{2}$, the RG beta functions are \cite{zhu2002}
\begin{eqnarray}
\beta(J) &=&-J \left[ (N_0 J) - \frac{1}{2} (N_0 J)^2
    - (K_0 g)^2 + (K_0 g)^4 \right], \nonumber \\
\beta(g) &=& -g \left[ \frac{\epsilon}{2}-(K_0g)^2 + (K_0g)^4  
	-\frac{1}{2} (N_0 J)^2 \right]. 
\label{rg-equations}
\end{eqnarray} 
Note that here and below, in Eq.\ \eqref{rg_ising}, 
 the field-theory sign convention (as opposed to the condensed-matter one) for the beta functions
has been used.

\begin{figure*}[t!]
\centering
  \mbox{\includegraphics[width=0.6\columnwidth]{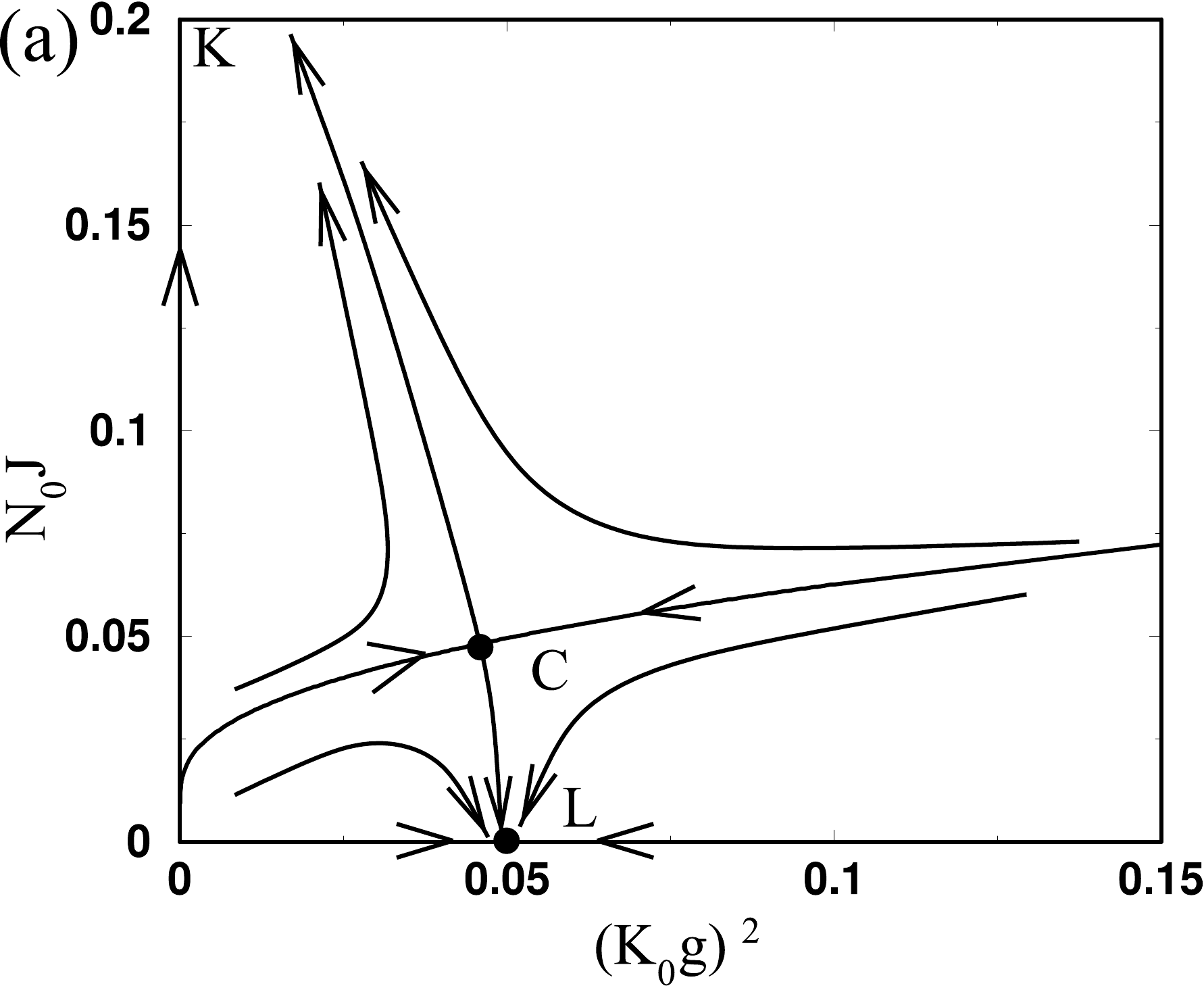}}    
  \mbox{\includegraphics[width=0.6\columnwidth]{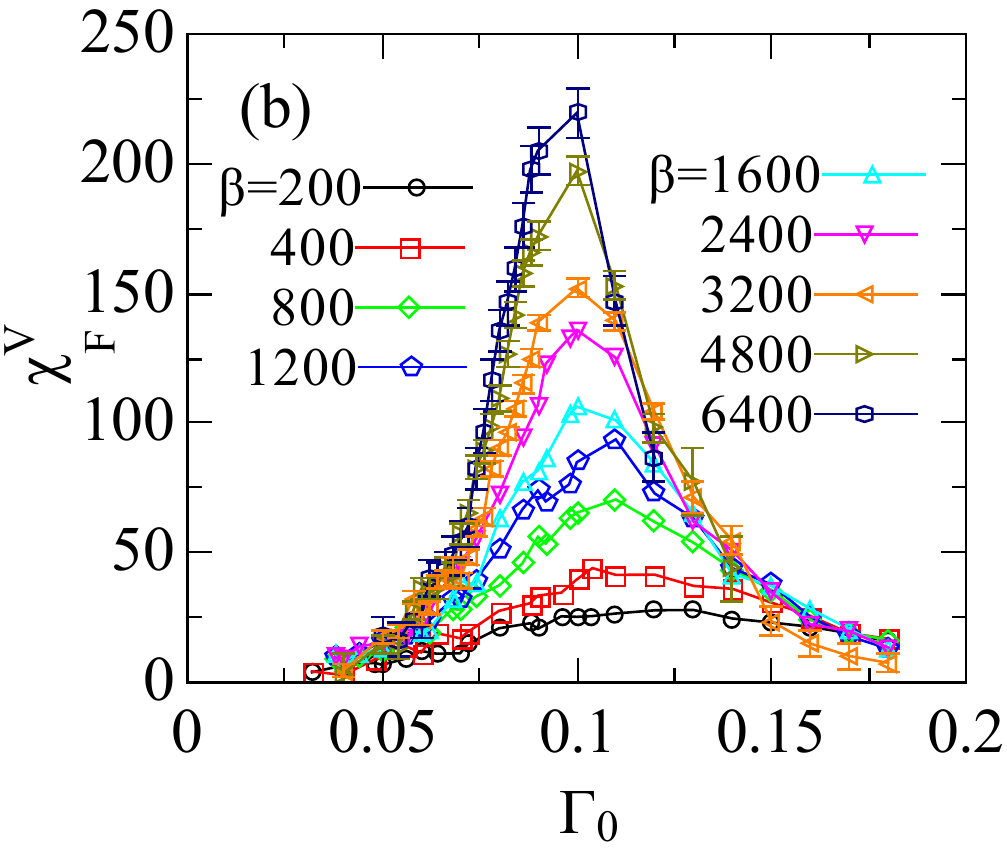}}    
  \mbox{\includegraphics[width=0.6\columnwidth]{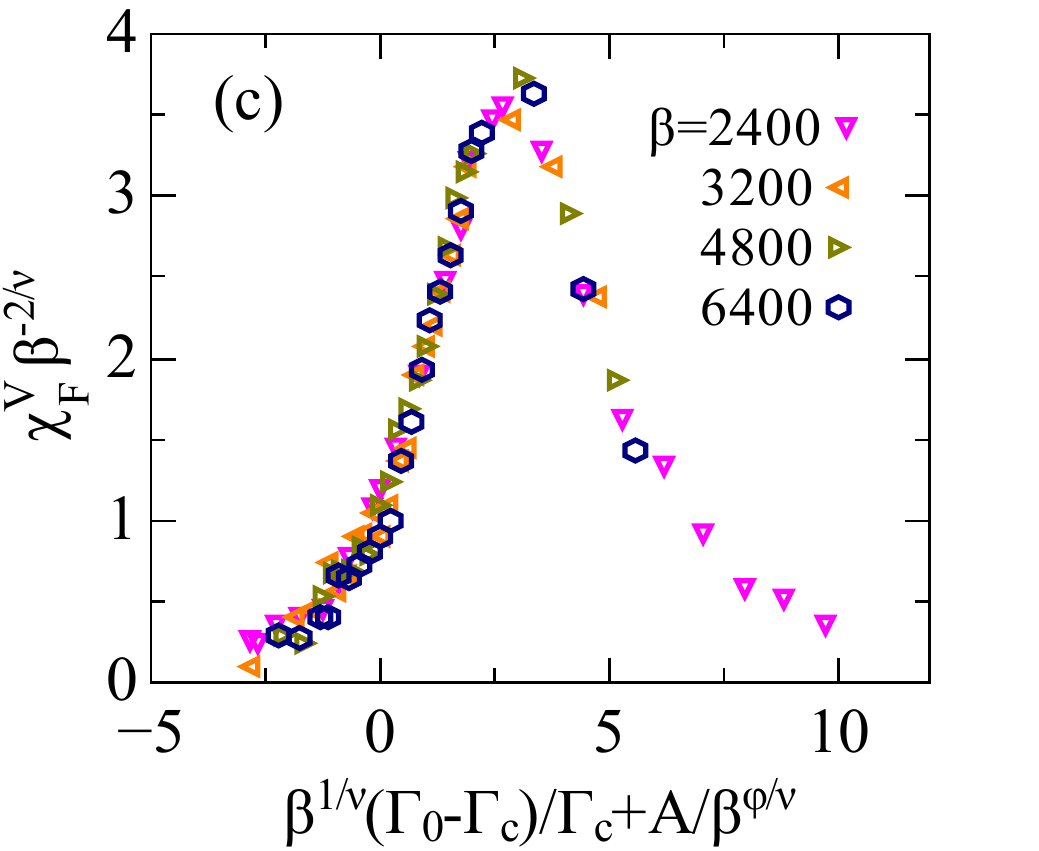}}    
\caption{\label{fig:chif_su2}
(a) RG flows on the $g$--$J$ plane for the SU(2)-symmetric BFK model with $\epsilon=1-s=0.1$,
adapted from Ref.\ \onlinecite{zhu2002}.
(b) Fidelity susceptibility vs $\Gamma_{0}$ for the SU(2)-symmetric BFA model with
$U=-2\epsilon_f=0.1$, $s=0.6$, and $g=0.5$ at inverse temperatures $\beta$ specified in the
legend.
(c) Finite-size scaling of the data in (b), plotted according to Eq.\ \eqref{eq:chif}.}
\end{figure*}

These RG equations yield two stable fixed points: the Kondo fixed point (K) at $g=0$ and large $J$,
and the local-moment fixed point (L) at $J=0$ and intermediate $g$. The equations also predict
two unstable fixed points: the noninteracting fixed point at $g=J=0$, and the critical point (C)
at nonzero values of $g$ and $J$ where Kondo destruction occurs.
Fig. \ref{fig:chif_su2}(a) illustrates the RG flows for the case $\epsilon=0.1$.

\subsubsection{CT-QMC treatment of the BFA model}

The SU(2)-symmetric BFA model can be solved using a recently developed extension of the CT-QMC
method \cite{cai2019,otsuki2013spin}. All CT-QMC results reported in this paper were obtained for
particle-hole-symmetric local level parameters $U=-2\epsilon_d=0.1$, for bosonic bath exponent
$s = 0.6$, and for fixed bosonic coupling $g=0.5$, using the hybridization width $\Gamma_{0}$
as the tuning parameter.

Within the CT-QMC approach, we can detect a quantum phase transition via a divergence of the
fidelity susceptibility $\chi^{\lambda}_{F}$. We consider the Hamiltonian to be composed of
two parts, $H=H_{0}+\lambda H_{1}$, where $\lambda$ is a real, dimensionless tuning parameter
(with $H_{0}$ and $H_{1}$ both being independent of $\lambda$). 
The fidelity between the ground states for parameters $\lambda$ and $\lambda+d\lambda$
is defined to be the modulus of the ground-state overlap \cite{you2007fidelity}:
\begin{eqnarray}
F(\lambda,d\lambda)
&=& \left|\langle \Psi_{0}(\lambda+d\lambda)|\Psi_{0}(\lambda) \rangle\right| \nonumber \\
&=& 1-\frac{(d\lambda)^2}{2} \chi^{\lambda}_{F}(\lambda) + O\left((d\lambda)^3\right),
\end{eqnarray}
where $\chi^{\lambda}_{F}$ is the zero-temperature fidelity susceptibility.

The definition of fidelity suceptibility can be extended to $T>0$ via \cite{albuquerque2010}
\begin{equation}
\chi^{\lambda}_{F}(T) = \int_{0}^{\beta/2} \left( \langle T_{\tau} H_{1}(\tau) H_{1} \rangle
  - \langle H_{1} \rangle^{2} \right) \tau \, d\tau,
\label{eq:fidel_original}
\end{equation}
where $\beta=1/T$ is the inverse temperature.
For a value of $\lambda$ that places the system at a QCP, $\chi^{\lambda}_{F}(T)$ will
diverge as \cite{albuquerque2010} $\chi^{\lambda}_{F}(T) \sim T^{-2/\nu}$ as $T\to 0$,
where $\nu$ is the correlation-length exponent. For all other values of $\lambda$,
$\chi^{\lambda}_{F}(T)$ saturates at a finite value for $T\to 0$.
Therefore, $\chi^{\lambda}_{F}(T)$ can be used to detect the location of the QCP.

Here, we choose the tuning parameter to be the hybridization $V$ (or strictly,
$\lambda = V/D$), and calculate $\chi_{f}^{V}$ vs $\Gamma_{0}$ at various temperatures
based on
the method
of Ref.\ \onlinecite{wang2015fidelity} 
and using the SU(2) CT-QMC approach of
Ref.\ \onlinecite{cai2019}.
As shown in Fig.\ \ref{fig:chif_su2}(b), the fidelity susceptibility develops a peak near
$\Gamma_{0}=0.1$ that becomes more pronounced as the temperature is lowered.

To confirm the asymptotic divergence of the fidelity susceptibility at the QCP, we can perform
finite-size scaling according to
\begin{equation}
\chi^{V}_{F}(\Gamma_{0},\beta) = \beta^{2/\nu}
  \tilde{\chi} \left(\beta^{1/\nu}(\Gamma_{0}-\Gamma_{c})/\Gamma_{c}+A/\beta^{\phi/\nu} \right).
\label{eq:chif}
\end{equation}
Here, $\Gamma_{c}$ is the critical value of $\Gamma_{0}$, while $A$ and $\phi$ parametrize
subleading corrections to the scaling \cite{beach2005data}. We optimize the choices of
$\Gamma_c$, $\nu^{-1}$, $A$, and $\phi$ by minimizing a ``quality function'' \cite{houdayer2004low}.
The collapsed data are plotted in Fig.\ \ref{fig:chif_su2}(c). We find $\Gamma_{c}=0.08(1)$
and $\nu^{-1}=0.24(4)$, consistent with the $\epsilon$-expansion result
$\nu^{-1}= \epsilon/2+\epsilon^{2}/6 +O(\epsilon^{3}) \simeq 0.23$.

\subsection{Ising anisotropy}

We now turn to Kondo destruction in Bose-Fermi models with Ising spin symmetry.

\subsubsection{$\epsilon$-expansion RG for the BFK model}

We again recall the RG analysis, which for Ising anisotropy can be carried out by
$\epsilon$-expansion in a kink-gas representation \cite{si1996kosterlitz,zhu2002}. 
The RG equations are given in terms of stiffness constants $\kappa_{j}$, $\kappa_{g}$ and
a fugacity $y_{j}$, which are related to the bare Hamiltonian parameters by
\begin{eqnarray}
\kappa_j &=& \big[1 - \pi^{-1} \tan^{-1}(\pi N_0 J_z / 4) \big]^2
\nonumber\\
\kappa_g &=&
{\textstyle\frac{1}{4}} \Gamma(\gamma) \tau_0^{1 - \gamma} (K_0g_z)^2
\nonumber\\
y_j &=& \half N_0 J_{\perp}.
\label{RG-charges-Ising}
\end{eqnarray}

The corresponding RG beta functions are given by\cite{si1996kosterlitz}
\begin{eqnarray}
\beta(\kappa_j)&=&4\kappa_{j}y_{j}^{2} \nonumber \\
\beta(\kappa_g)&=&-\kappa_{g}(\epsilon-4 y_{j}^{2}) \nonumber \\
\beta(y_{j})&=&-y_{j}(1-\kappa_{j}-\kappa_{g}/2).
\label{rg_ising}
\end{eqnarray}

The stiffness $\kappa_j$ is irrelevant, so all fixed points occur at $\kappa_j=0$.
RG flows on the $\kappa_{g}$--$y_{j}$ plane are shown schematically in
Fig.\ \ref{fig:binder}(a). There are two stable fixed points: the Kondo fixed
point (K) at $\kappa_{g}=0$ and large $y_{j}$, and the local-moment fixed point (L) at
$y_{j}=0$ and large $\kappa_{g}$. There are also two unstable fixed points: the
noninteracting point at $\kappa_{g}=y_{j}=0$, and the critical point (C) at nonzero
values of $\kappa_g$ and $y_j$, which is the Kondo-destruction critical point.

\subsubsection{CT-QMC calculations for the BFA model}

Our CT-QMC calculations for the Ising-anisotropic BFA model were carried out along the same
lines as ones reported previously \cite{pixley2013quantum} for 
such a model
with a
pseudogapped fermionic density of states. The results reported in 
the present
paper were
obtained using the same fixed parameters $U=-2\epsilon_f=0.1$, $s = 0.6$, and $g=0.5$ as
in the SU(2)-symmetric case.

A quantity that can be used to demonstrate the existence of a second-order phase transition
in CT-QMC is the Binder ratio \cite{binder1981finite}
\begin{equation}
B= \frac{ \langle m_{z}^{4} \rangle }{ \langle m_{z}^{2} \rangle^{2} }  ,
\end{equation}
where $m_{z} = \frac{1}{\beta} \int_{0}^{\beta} {S} _ {z} (\tau) \, d\tau$ is the
local magnetization.

The Binder ratio has limiting values $B= 3$ deep in the Kondo-screened phase, where the
distribution of $m_{z}$ is Gaussian centered around $0$, and $B= 1$ deep in the 
Kondo-destroyed phase where the distribution of $m_{z}$ is two delta functions centered around
$\pm 1/2$. Furthermore, $B$ is defined in a way such that at a QCP, it should have a scaling
dimension exactly equal to $0$, thereby becoming independent of temperature. Thus, a QCP can
be identified from data at $T>0$ ($\beta<\infty$) as the crossing point of $B$ vs $\Gamma_{0}$
curves for different values of $\beta$.

\begin{figure*}[t!]
\centering
  \mbox{\includegraphics[width=0.48\columnwidth]{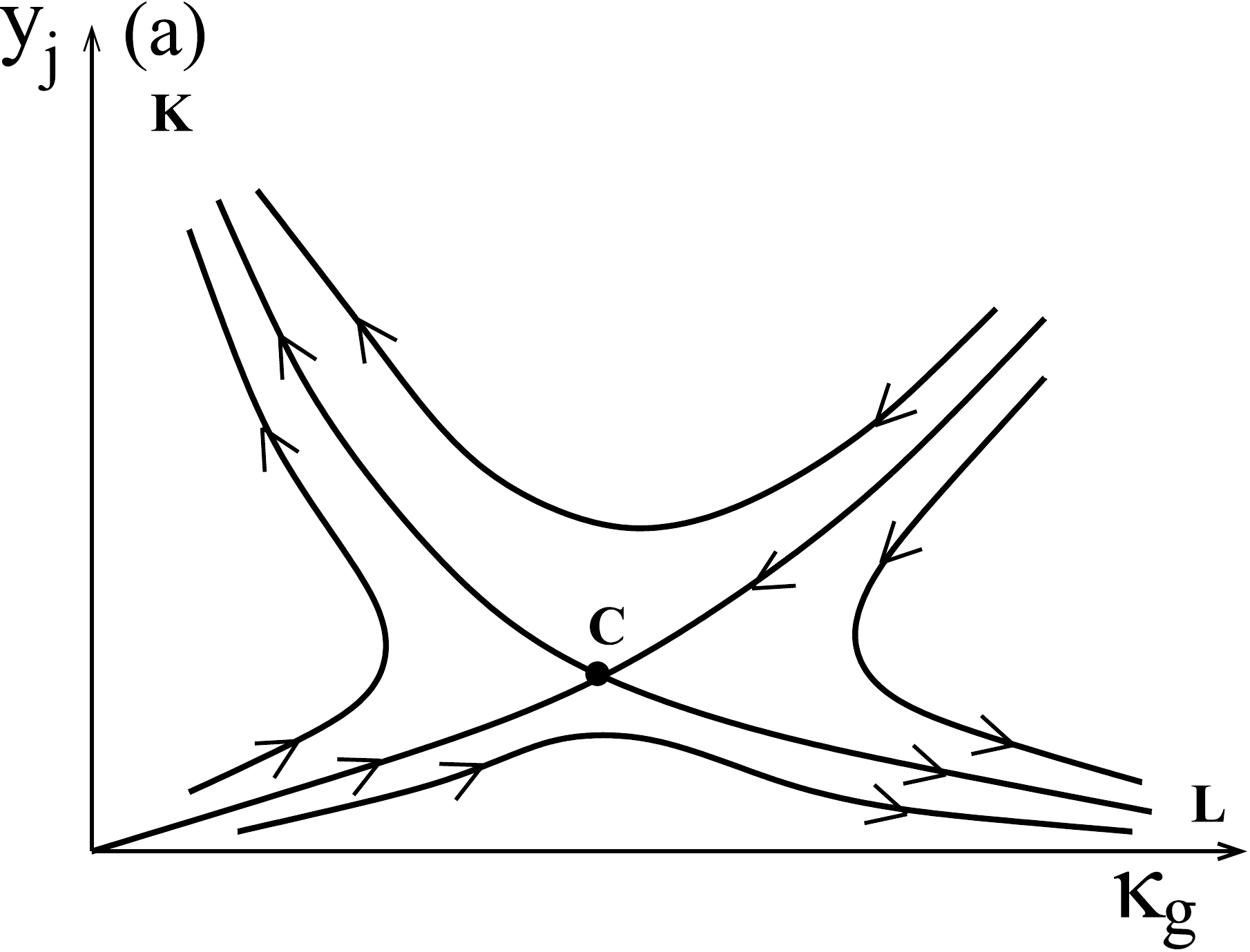}}    
  \mbox{\includegraphics[width=0.48\columnwidth]{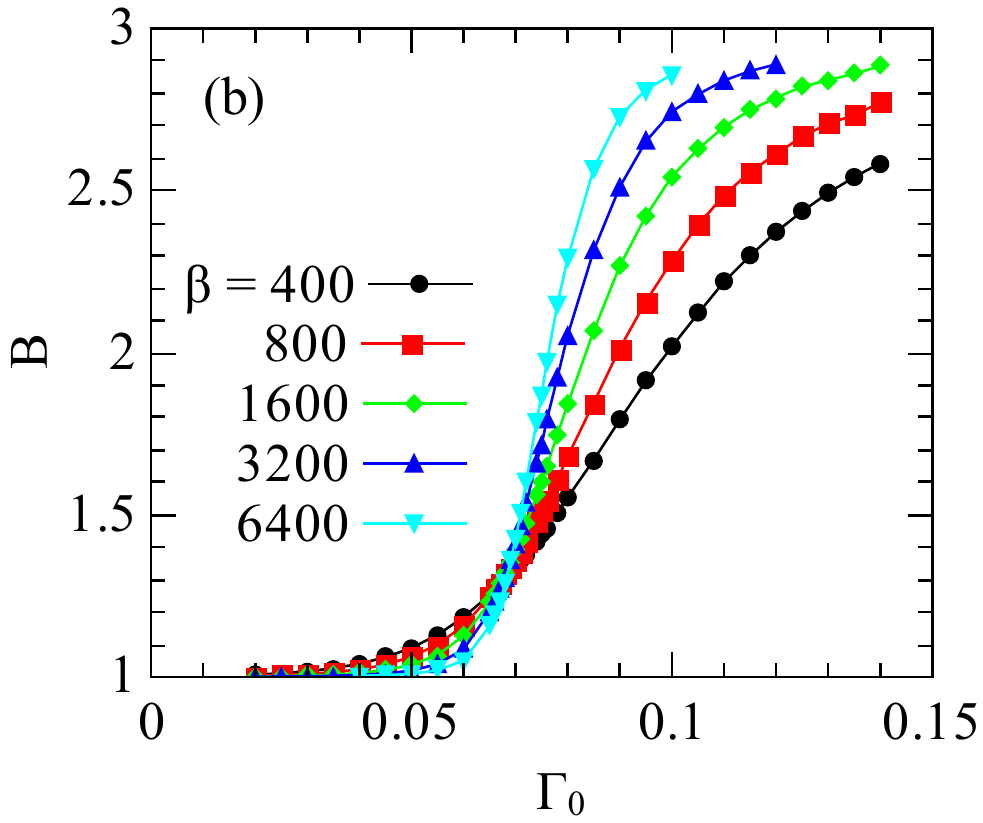}}    
  \mbox{\includegraphics[width=0.48\columnwidth]{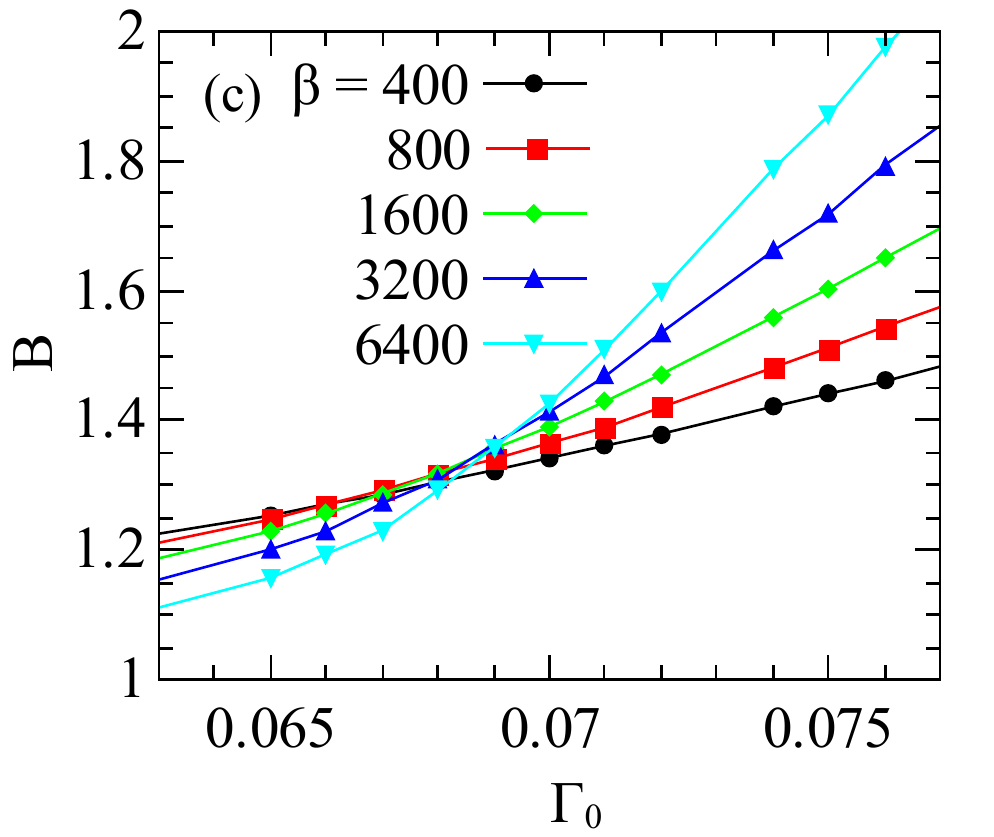}}    
  \mbox{\includegraphics[width=0.54\columnwidth]{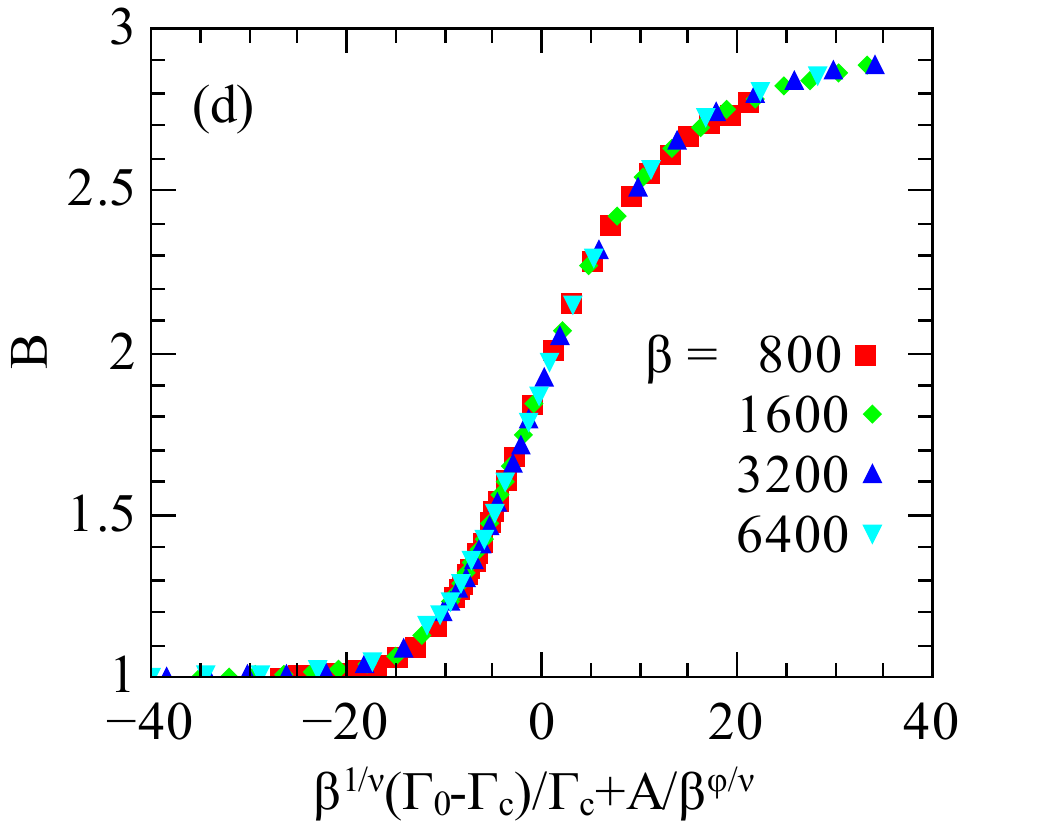}}    
\caption{\label{fig:binder}
(a) Schematic RG flow diagram for the Ising-anisotropic BFK model, adapted from Ref.\
\onlinecite{zhu2002}.
(b) Binder ratio $B$ vs $\Gamma_{0}$ for the Ising-anisotropic BFAM with
$U=-2\epsilon_f=0.1$, $s=0.6$, and $g=0.5$ at inverse temperatures $\beta$ specified in the
legend.
(c) Zoom-in of $B$ vs $\Gamma_{0}$ near the crossing point.
(d) Finite-size scaling of the Binder ratio in (c), plotted according to Eq.\ \eqref{eq:Binder}.
These results provide evidence for the existence of a QCP and a Kondo-destroyed phase, consistent
with the analytical prediction in (a) from the $\epsilon$-expansion RG.}
\end{figure*}

Numerical data for $B$ vs $\Gamma_{0}$ are shown in Fig.\ \ref{fig:binder}(b). A zoomed-in view in
Fig.\ \ref{fig:binder}(c) clearly shows a crossing around $\Gamma_{0}=0.07$, suggesting the existence
of a QCP between a Kondo-destroyed phase at small $\Gamma_{0}$ (where $B \rightarrow 1$) and a
Kondo-screened phase at large $\Gamma_{0}$ (where $B \rightarrow 3$).
To support this interpretation, we can perform finite-size scaling according to \cite{beach2005data}
\begin{eqnarray}
B(\Gamma_{0},\beta)
  = \tilde{U}_{2} \left(\beta^{1/\nu}(\Gamma_{0}-\Gamma_{c})/\Gamma_{c}+A/\beta^{\phi/\nu} \right) .
\label{eq:Binder}  
\end{eqnarray}
The optimal data collapse, plotted in Fig.\ \ref{fig:binder}(d), occurs for $\Gamma_{c}=0.07(1)$
and $\nu^{-1}=0.51(4)$. The latter value is consistent with a previous NRG study of the corresponding BFK
model \cite{pixley2013quantum}, where $\nu^{-1}=0.509(1)$ was reported.

\begin{figure}[htb!]
\captionsetup[subfigure]{labelformat=empty}
  \centering
    \mbox{\includegraphics[width=0.7\columnwidth]{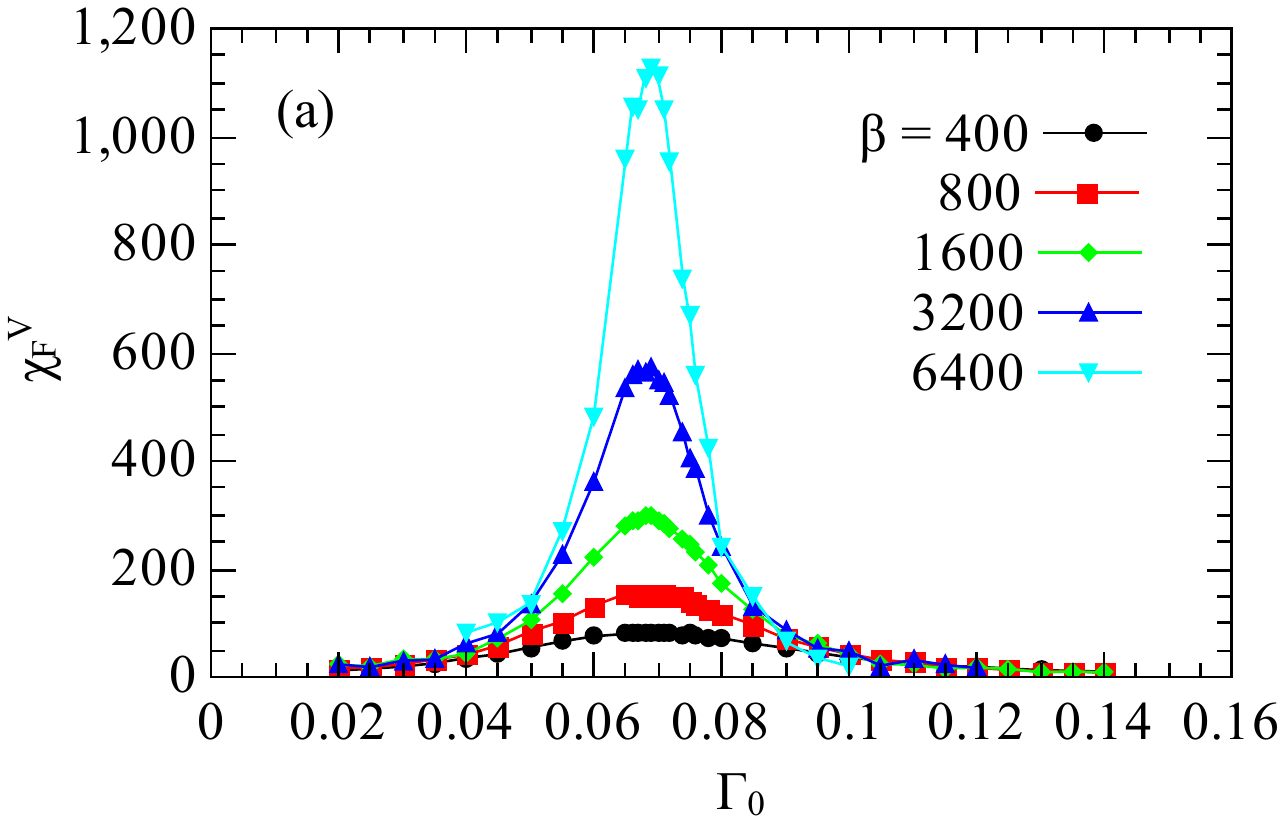}}    
  \mbox{\includegraphics[width=0.7\columnwidth]{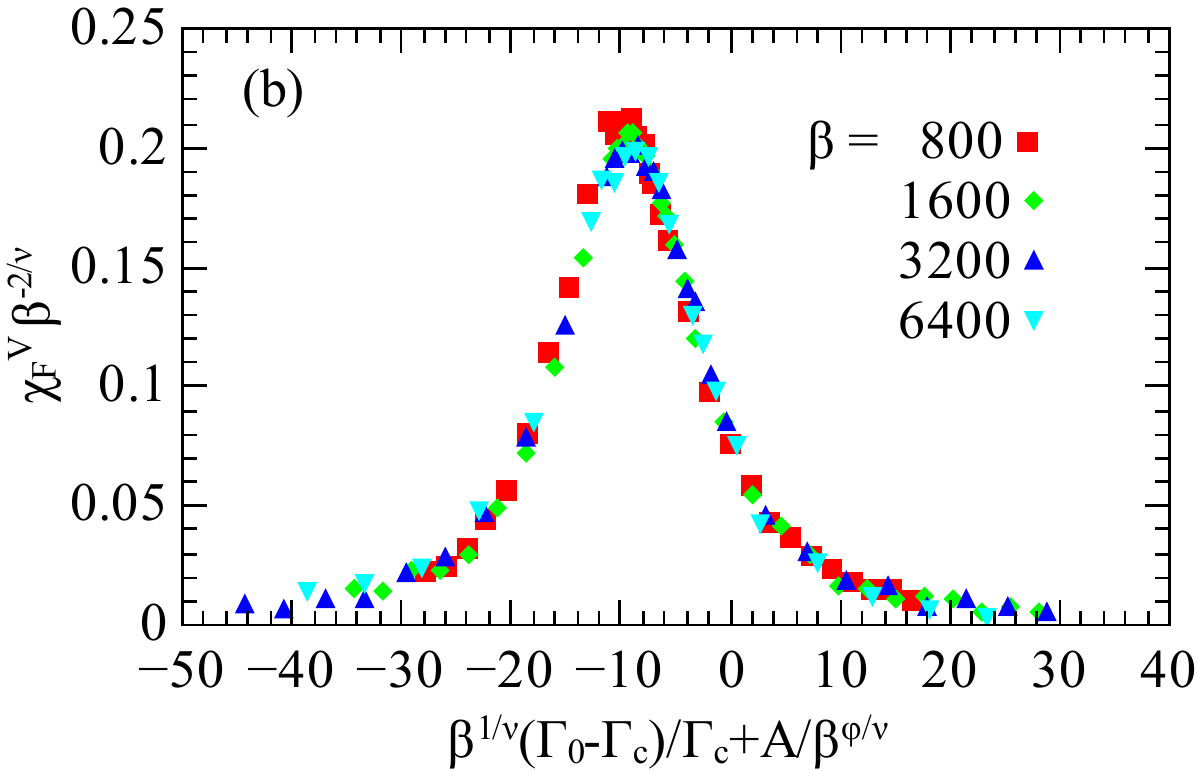}}    
\caption{\label{fig:chif_ising}
(a) Fidelity susceptibility vs $\Gamma_{0}$ for the Ising-anisotropic BFAM with $U=-2\epsilon_f=0.1$,
$s=0.6$, and $g=0.5$ at inverse temperatures $\beta$ specified in the legend.
(b) Finite size-scaling of the data in (a), demonstrating the divergence of the fidelity susceptibility
and the associated scaling collapse, both providing evidence for the existence of a QCP. These results
provide a consistency check for the QCP and Kondo-destroyed phase evidenced by the crossing and scaling
of the Binder ratio in Fig.\ \ref{fig:binder}.}
\end{figure}

We also confirm the existence of a Kondo-destruction QCP by calculating the fidelity susceptibility
$\chi_{F}^{V}$ vs $\Gamma_{0}$ at various temperatures, revealing a peak near $\Gamma_{0}=0.07$; see
Fig.\ \ref{fig:chif_ising}(a). When scaled according to Eq.\ \eqref{eq:chif}, the data collapse as
shown in Fig.\ \ref{fig:chif_ising}(b). The scaling gives $\Gamma_{c}=0.07(1)$ and $\nu^{-1}=0.50(4)$,
in agreement with the results obtained from the Binder ratio.

\subsubsection{NRG study of the BFA model}

In this subsection, we present results for the Ising BFA model obtained using the Bose-Fermi NRG
\cite{glossop2005bfk,glossop2007bfk}, which complements the CT-QMC by providing direct access to $T=0$
properties. The NRG discretization parameter $\Lambda$ was chosen to be $\Lambda=9$, the bosonic
occupation of each Wilson chain site was capped at 8, and up to 500 many-body charge (``isospin'')
multiplets were retained after each NRG iteration.
We used the same bosonic bath exponent $s=0.6$ as in the CT-QMC calculations, and worked with fixed
Hamiltonian couplings $U=-2\epsilon_{d}=0.2$ and $K_{0}g\simeq 1.026$ such that the QCP is located at
$\Gamma_{0}=\Gamma_{c}=0.1$. 

\begin{figure}[htb!]
\captionsetup[subfigure]{labelformat=empty}
  \centering
    \mbox{\includegraphics[width=0.7\columnwidth]{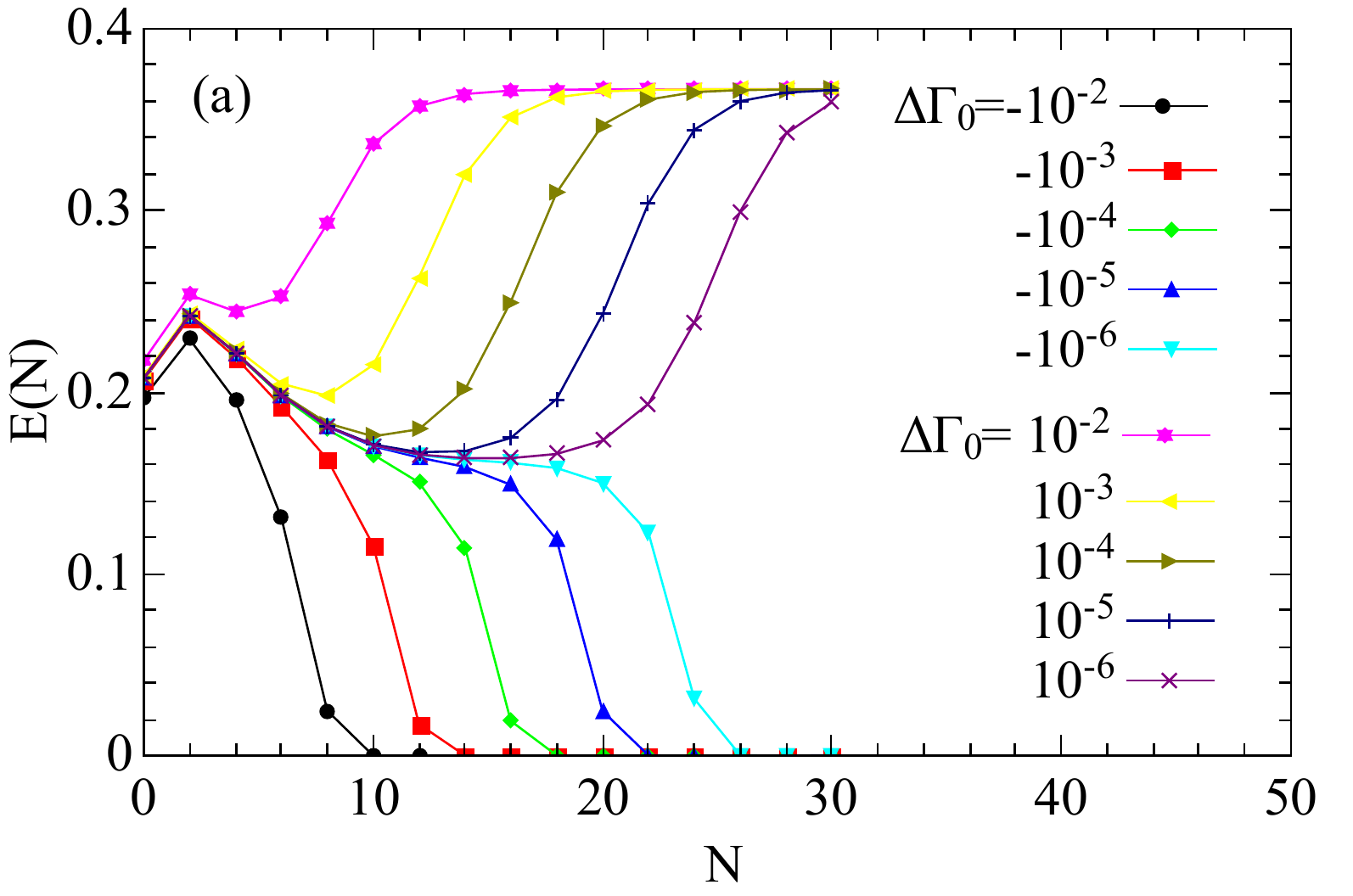}}   
    \mbox{\includegraphics[width=0.68\columnwidth]{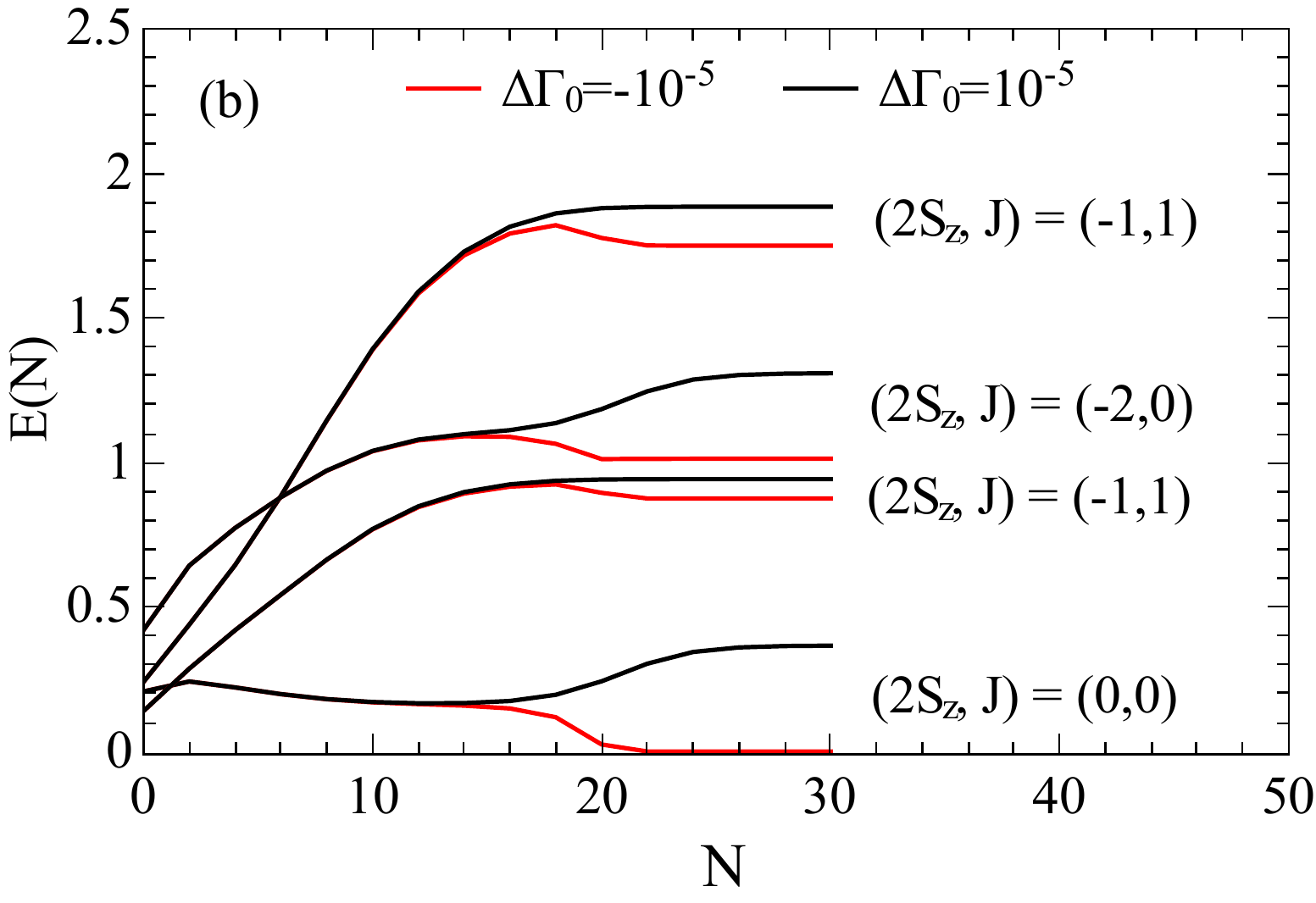}}   
     \mbox{\includegraphics[width=0.85\columnwidth]{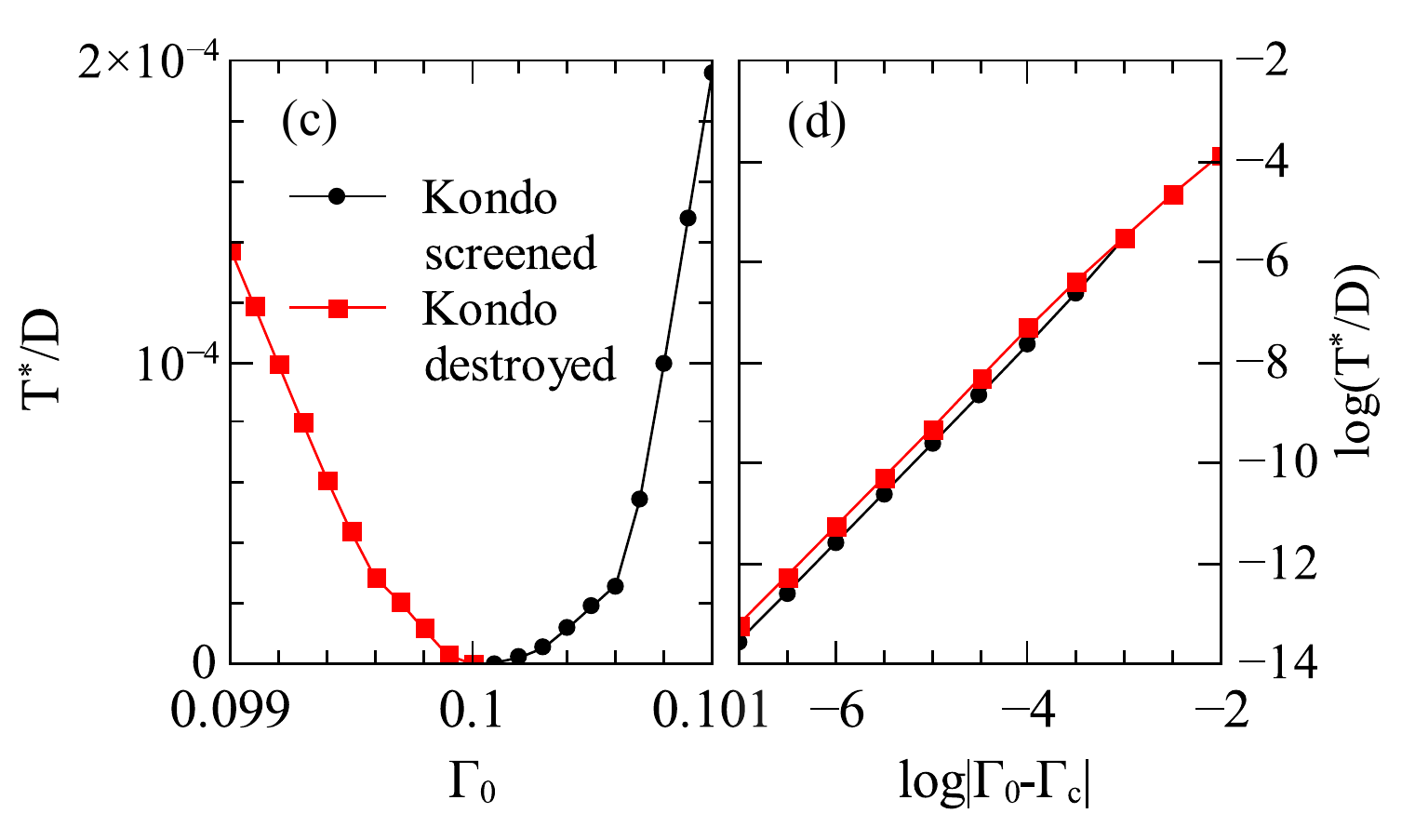}}    
\caption{\label{fig:nrg_flow}
NRG data for the Ising-anisotropic BFA model with $U=-2\epsilon_f=0.2$, $s = 0.6$,
and $K_{0}g\simeq 1.026$.
(a) Flow of the lowest scaled excited-state energy vs even iteration number $N$ for
different values $\Gamma_{0}-\Gamma_c=\pm 10^{-p}$ with $p=2,3,\dots,6$.
(b) Several scaled excitation energies $E(N)$ vs even iteration number $N$ for $\Gamma_{0}$
slightly below and above the critical coupling $\Gamma_{c}=0.1$,
labeled by their quantum numbers $S_{z}$ ($z$ component of total spin) and $J$ (total isospin). 
(c) Crossover temperature $T^{*}$ vs $\Gamma_{0}$ over a broad range on a
linear plot, demonstrating the vanishing of $T^{*}$ at the QCP at $\Gamma_{0}=0.1$.
(d) Power-law behavior of the crossover scale $T^{*}$ in the vicinity of the QCP,
showing data in the Kondo-screened phase (red squares) and the Kondo-destroyed
phase (black circles).}
\end{figure}

The existence of distinct phases and quantum phase transitions can be seen from the flow
of the NRG many-body spectrum as a function of the iteration number $N$, as illustrated in
Fig.\ \ref{fig:nrg_flow}. The dependent quantity $E$ is a many-body eigenenergy at iteration
$N$ divided by $\half\Lambda^{1/2}(1+\Lambda^{-1}) D\Lambda^{-N/2}$, the characteristic energy
scale of site $N$ on the Wilson fermionic chain. With this rescaling, an RG fixed point is
associated with a scale invariant spectrum where each eigenenergy satisfies $E(N+2)=E(N)$.
Fig. \ref{fig:nrg_flow}(a) shows the NRG flow of the lowest excited state for even iterations
at different distances $\Delta \Gamma_{0}=\Gamma_{0}-\Gamma_{c}$ from the QCP, specifically,
$\Delta \Gamma_{0}=\pm 10^{-p}$ with $p=2,3,\dots,6$. The bifurcation of the energy flows
for large $N$ clearly demonstrates the existence of a QCP. For values of $\Gamma_0$ close to
$\Gamma_c$, the scaled energy $E(N)$ quickly converges over the first dozen or so iterations
to a value $E\simeq 0.167$ that forms part of the critical spectrum. In the Kondo phase
(i.e., for $\Gamma_{0}>\Gamma_{c}$), $E(N)$ rises for larger $N$ toward an asymptotic value
around $0.366$, corresponding to the energy for a free bosonic excitation, whereas in the
Kondo-destroyed phase ($\Gamma_{0}<\Gamma_{c}$), $E(N)$ descends toward zero.

Fig. \ref{fig:nrg_flow}(b) plots the flow of several low-lying excited states for
hybridization widths $\Gamma_0$ slightly below ($\Delta \Gamma_{0}=-10^{-5} $) and above
($\Delta\Gamma_{0}=10^{-5}$) the critical value $\Gamma_{c}=0.1$. The many-body eigenstates
are labeled by their values of $S_{z}$ (the $z$ component of the total spin) and $J$ (the total
isospin, an SU(2) quantum number whose $z$ component represents the total fermionic occupancy
measured from half-filling). Intermediate iterations $N \simeq 14$ access the quantum-critical
spectrum prior to flow toward either the Kondo or the local-moment fixed-point structure at
iterations $N\gtrsim 20$.

The flow of the bosonic energy levels shown in Fig.\  \ref{fig:nrg_flow}(a) can be used to
define a crossover scale $T^{*}=\half\Lambda^{1/2}(1+\Lambda^{-1}) D\Lambda^{-N^*/2}$ in each
phase. Here, $N^*$ is the interpolated iteration number (not necessarily an integer) at which
the scaled excitation energy $E$ passes a predetermined threshold value lying between the
critical and stable fixed-point excitation energies. Fig. \ref{fig:nrg_flow}(c) shows
$T^{*}$ vs $\Gamma_{0}$ on linear scales, demonstrating that $T^{*}$ vanishes on approach to the
QCP from either side. This provides direct evidence for Kondo destruction. The log-log plot
of $T^{*}$ vs $|\Gamma_{0}-\Gamma_{c}|$ in Fig.\ \ref{fig:nrg_flow}(d) demonstrates that
$T^{*}$ vanishes in a power-law fashion. Since this energy scale characterizes the crossover
from the quantum-critical regime to the Fermi-liquid regime near a stable fixed point, it is
a measure of the quasiparticle weight $Z$. The NRG data thus establish that $Z$ goes continuously
to zero as the QCP is approached from {\it either} side.

\section{Dynamical Kondo effect}
\label{sec:sdots}

We now turn to the calculation of $\langle {\bf S}_{f} \cdot {\bf s}_{c} \rangle $ and
$\langle S_{f}^{z} s_{c}^{z}\rangle$.
We show that these expectations values are nonzero in
the Kondo-destroyed phase, and lay out the implications of this observation for the dynamical
Kondo effect.

\subsection{Nonzero $\langle {\bf S}_{f} \cdot {\bf s}_{c} \rangle $ and
$\langle S_{f}^{z} s_{c}^{z} \rangle$ in the Kondo-destroyed phase}

\begin{figure}[htb!]
\captionsetup[subfigure]{labelformat=empty}
  \centering
    \mbox{\includegraphics[width=0.7\columnwidth]{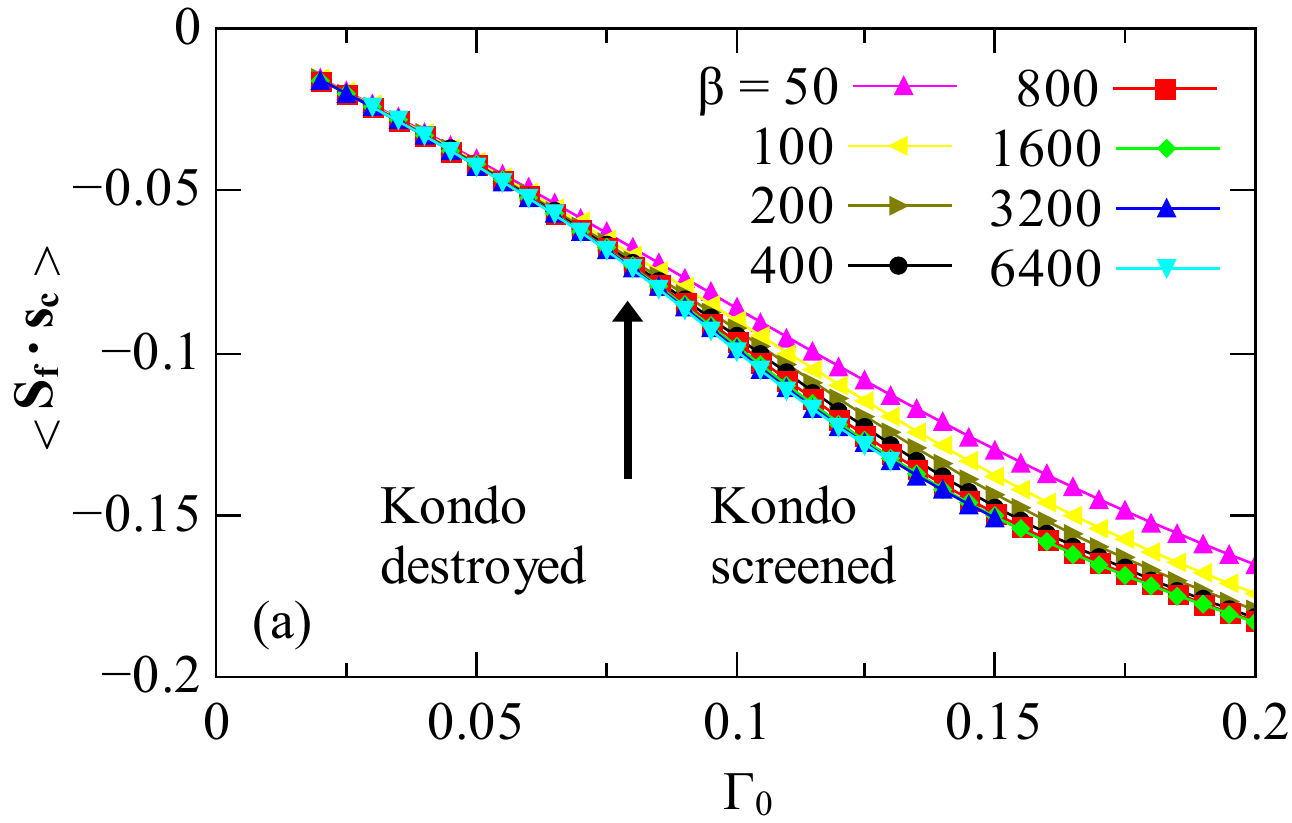}}    
     \mbox{\includegraphics[width=0.7\columnwidth]{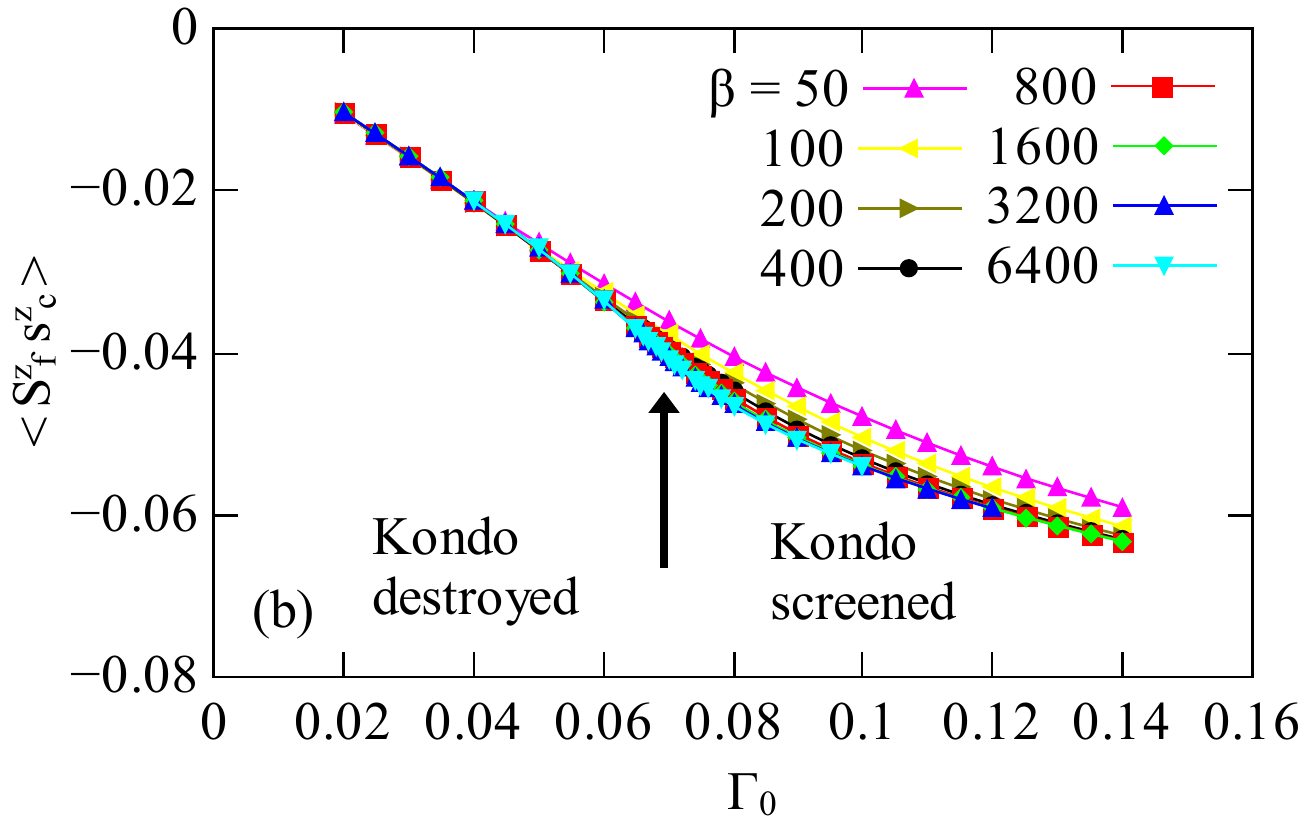}}   
     \mbox{\includegraphics[width=0.7\columnwidth]{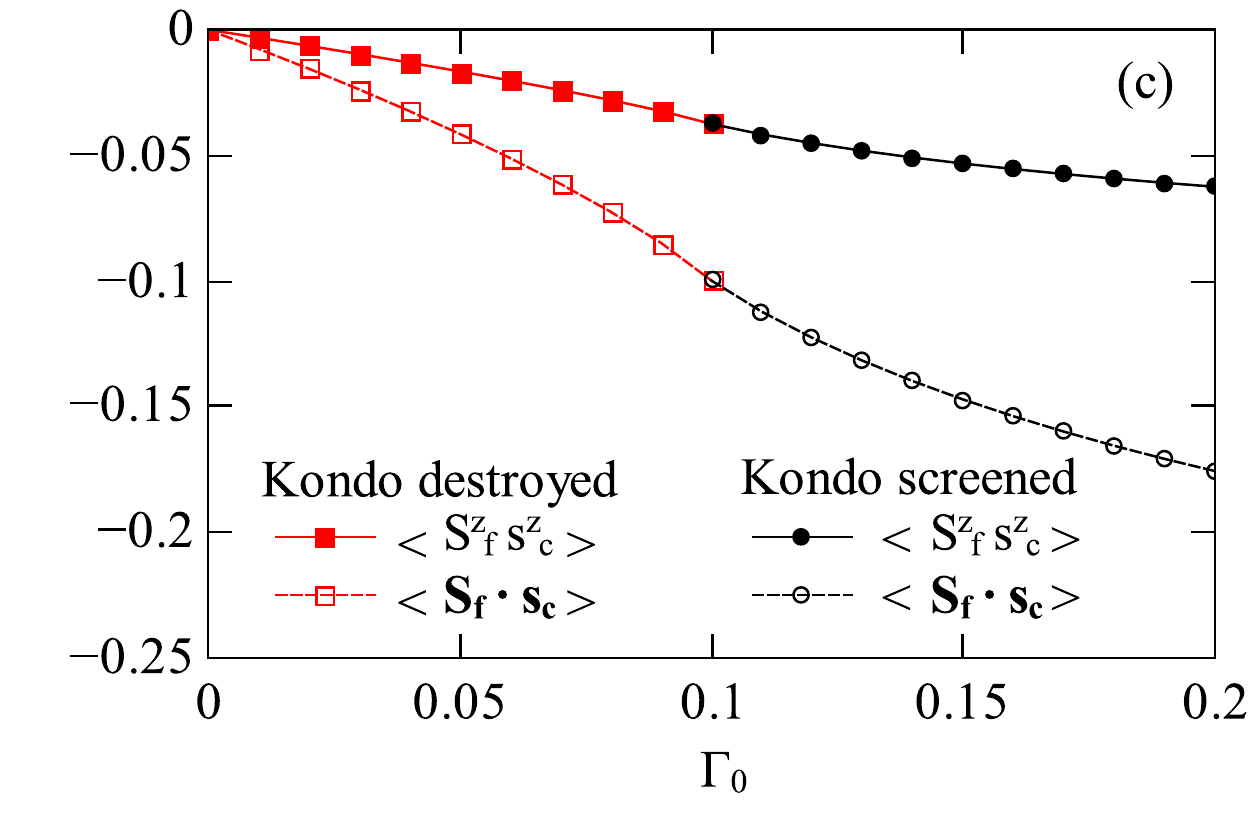}}   
\caption{\label{fig:sfsc}
(a) CT-QMC data for $\langle {\bf S}_{f} \cdot {\bf s}_{c} \rangle $ vs $\Gamma_{0}$ in the
SU(2)-symmetric BFA model with $U=-2\epsilon_f=0.1$, $s=0.6$, and $g=0.5$ at inverse
temperatures $\beta$ specified in the legend. The arrow marks the QCP at $\Gamma_0=\Gamma_c$,
which separates the Kondo-destroyed and Kondo-screened phases.
(b) Corresponding plot of $\langle S_{f}^{z} s_{c}^{z} \rangle $ vs $\Gamma_{0}$ in the
Ising-anisotropic BFA model.
(c) NRG results at $T=0$ for $\langle {S}_{f}^{z} {s}_{c}^{z} \rangle $ (solid symbols) and
$\langle {\bf S}_{f} \cdot {\bf s}_{c} \rangle$ (open symbols) vs $\Gamma_0$ in the Ising BFAM
with $U=-2\epsilon_f=0.2$, $s = 0.6$, and $K_{0}g\simeq 1.026$. Red and black symbols plot
values, respectively, on the Kondo-destroyed and Kondo-screened sides of the QCP.}
\end{figure}

Fig. \ref{fig:sfsc}(a) and Fig. \ref{fig:sfsc}(b) show CT-QMC results for, respectively,
$\langle {\bf S}_{f} \cdot {\bf s}_{c} \rangle $ in the SU(2)-symmetric BFA model and
$\langle S_{f}^{z} s_{c}^{z} \rangle $ in the Ising-anisotropic BFA model.
The calculation were performed using the same sets of parameters as in Sec.\ \ref{sec:existence},
where a Kondo-destruction QCP was identified in each model. One sees that
$\langle {\bf S}_{f} \cdot {\bf s}_{c} \rangle $ and $\langle S_{f}^{z} s_{c}^{z} \rangle$ are
continuous on passage through their respective QCPs (located at $\Gamma_{0}=0.08$ in the SU(2)
case and $\Gamma_{0}=0.07$ in the Ising case) and remain finite in the Kondo-destroyed phase,
only approaching $0$ when the hybridization parameter $\Gamma_{0}$ goes to zero. The data show
only weak temperature dependence.

Zero-temperature NRG results for both $\langle S_{f}^{z} s_{c}^{z} \rangle$ and
$\langle {\bf S}_{f} \cdot {\bf s}_{c} \rangle$ in the Ising-anistropic BFA model are plotted
in Fig.\ \ref{fig:sfsc}(c).
These expectation values are continuous across the QCP at $\Gamma_{0}= \Gamma_{c} = 0.1$,
consistent with the CT-QMC results obtained for $T>0$. Repeating the NRG calculations for
different choices of $U$ and $\epsilon_{f}$ reveals that $\langle S_{f}^{z} s_{c}^{z} \rangle$
and $\langle {\bf S}_{f} \cdot {\bf s}_{c} \rangle$ do not assume universal critical values, but
rather vary according to where one crosses the phase boundary between the Kondo-screened and
Kondo-destroyed phases.

Fig. \ref{fig:sfsc} represents the central result of our work: the expectation values 
$\langle {\bf S}_{f} \cdot  {\bf s}_{c} \rangle $  and $\langle S_{f}^{z} s_{c}^{z} \rangle$
are nonzero in the Kondo-destroyed phase even though the localized ($f$) spin is
asymptotically decoupled from the conduction band. As argued in Sec.\ \ref{sec:introduction}
based on Eq.\ \eqref{expectation_value_kk} and its analog for $\langle S_{f}^{z} s_{c}^{z} \rangle$,
these nonzero expectation values imply that the spectral functions 
$\operatorname{Im} \chi_{Ss}(\omega)$ and $\operatorname{Im} \chi^z_{Ss}(\omega)$ are nonzero at
frequencies $\omega\ne 0$, signifying the presence of the dynamical Kondo effect in the
Kondo-destroyed phase.
 
\subsection{Signature of the transition}

While the expectation values $\langle {\bf S}_{f} \cdot {\bf s}_{c} \rangle $ and
$\langle S_{f}^{z} s_{c}^{z} \rangle$ are continuous under variation of $\Gamma_0$
through $\Gamma_c$, the location of the Kondo-destruction QCP is revealed by the
the derivatives of these expectation values with respect to the control parameter.
Specifically, we focus on the susceptibilities
\begin{eqnarray}
\chi_{K}&=&-\frac{\partial \langle {\bf S}_{f} \cdot {\bf s}_{c} \rangle}{\partial \Gamma_{0}}, \\
\chi_{K}^{z}&=&-\frac{\partial \langle {S}^{z}_{f}  {s}^{z}_{c} \rangle}{\partial \Gamma_{0}}.
\label{eq:chizfc}
\end{eqnarray}

\begin{figure}[htb!]
\captionsetup[subfigure]{labelformat=empty}
  \centering
  \mbox{\includegraphics[width=0.66\columnwidth]{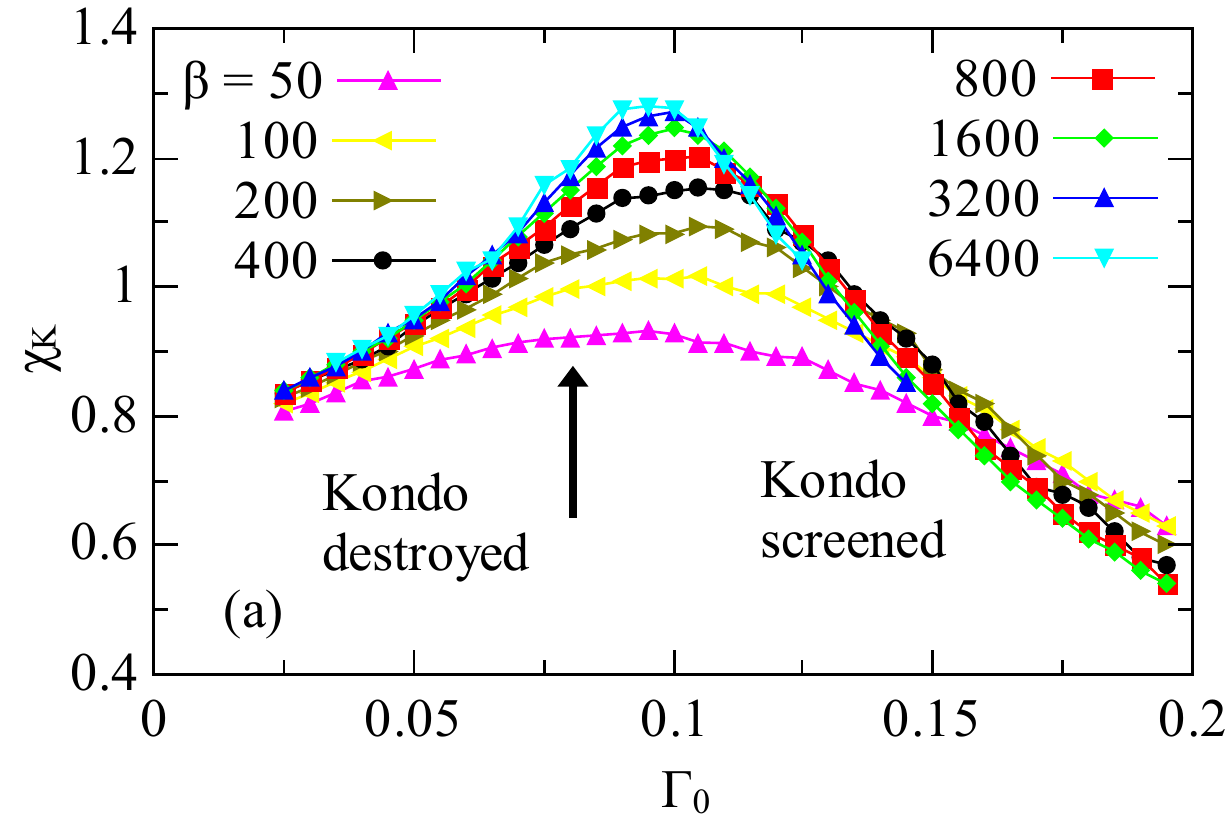}}    
  \mbox{\includegraphics[width=0.66\columnwidth]{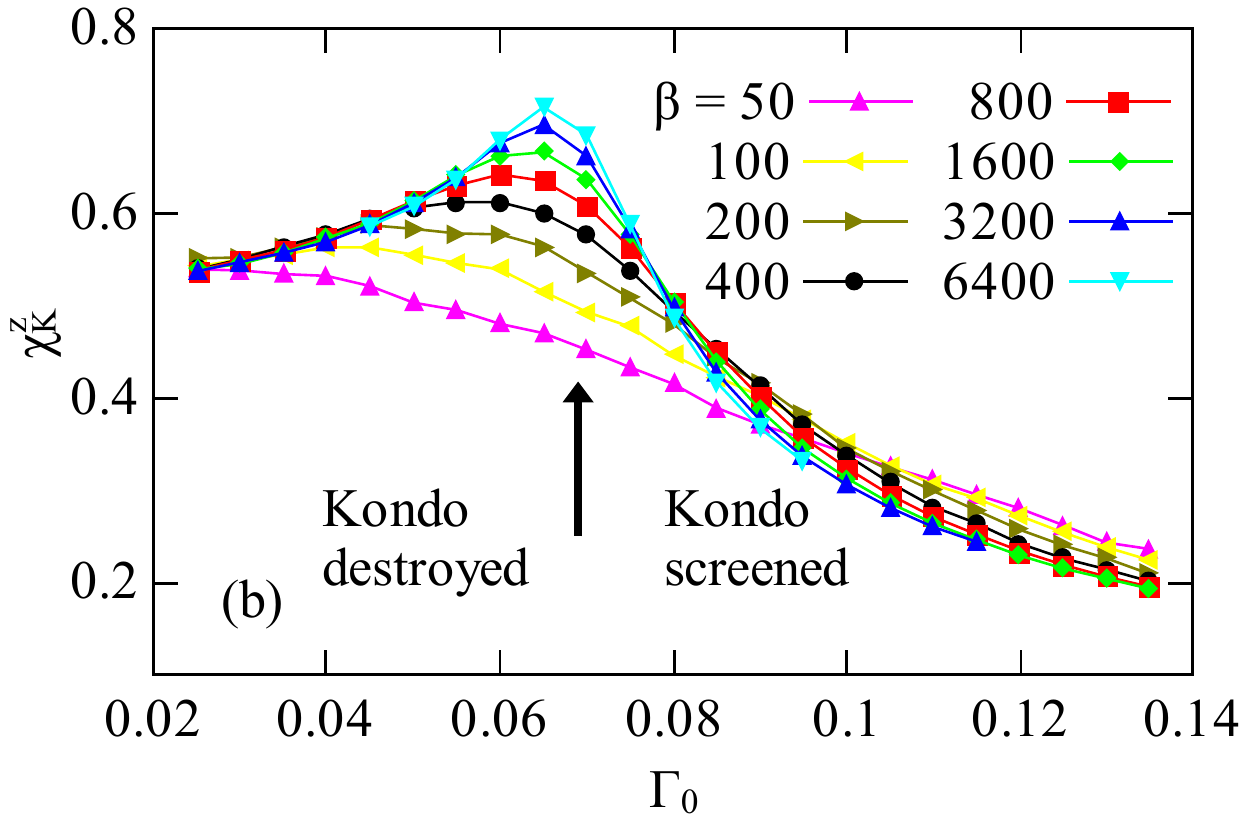}} 
   \mbox{\includegraphics[width=0.7\columnwidth]{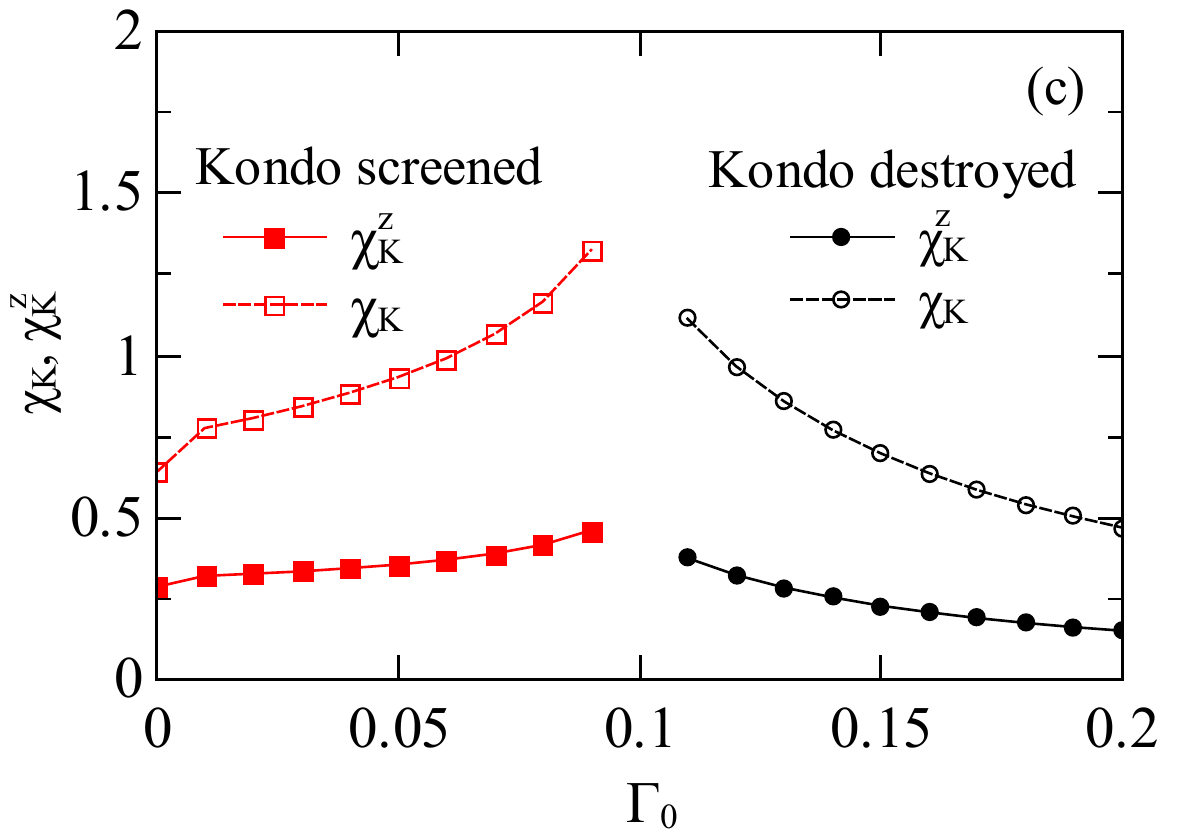}}    
\caption{\label{fig:xfc}
(a) Susceptibility $\chi_{K}$ vs $\Gamma_{0}$ for the SU(2)-symmetric BFA model with
$U=-2\epsilon_f=0.1$, $s=0.6$, and $g=0.5$ at inverse temperatures $\beta$ specified in the legend.
(b) Corresponding plot of $\chi^{z}_{K}$ vs $\Gamma_{0}$ for the Ising-anisotropic BFA model.
(c) NRG data for $\chi^{z}_{fc}$ and $\chi_{fc}$ vs $\Gamma_0$ in the Ising-anistropic BFAM
with $U=-2\epsilon_f=0.2$, $s = 0.6$, and $K_{0}g\simeq 1.026$.}
\end{figure}

Fig. \ref{fig:xfc}(a) and Fig. \ref{fig:xfc}(b) plot, respectively, CT-QMC results for $\chi_{K}$
in the SU(2)-symmetric BFA model and $\chi_{K}^{z}$ in the Ising-anisotropic BFA model. 
Both susceptibilities show peaks centered near $\Gamma_0 = \Gamma_c$ that gradually become more
pronounced as the temperature is lowered.

The corresponding $T=0$ NRG at for the Isinig-anisotropic BFA model is shown in Fig.\ \ref{fig:xfc}(c).
We see that $\chi_{K}^{z}$ and $\chi_{K}$ are increasing as we approach the QCP from both sides. 
This suggests that they will be peaked at the QCP, in agreement with the trend with lowering
temperatures shown in the CT-QMC result.

\section{Discussion}
\label{sec:discussion}

Several remarks are in order. First, we have shown that $\langle {\bf S}_{f} \cdot {\bf s}_{c}\rangle $
is nonzero in the Kondo-destroyed phase of two BFA models. The same property can be argued to
apply to BFK models. Consider, for definiteness, the SU(2)-symmetric BFK Hamiltonian
[Eq.\ \eqref{H-BFK-SU2}], which contains a term $J \, {\bf S}_{f} \cdot {\bf s}_c$. We can view 
$J$ as an external source field, with ${\bf S}_{f} \cdot {\bf s}_c$ being its conjugate.
This is in analogy to the case of a Zeeman coupling, where the magnetization is the conjugate
to an external magnetic field.
Similar to application of an external magnetic field leading to a nonzero magnetization,
we expect $\langle {\bf S}_{f} \cdot {\bf s}_{c} \rangle$ to be generically 
nonzero in the presence of a nonzero $J$, irrespective of whether the system is in its
Kondo-screened or Kondo-destroyed phase.

Second, that $\langle {\bf S}_{f} \cdot {\bf s}_{c} \rangle$ is continuous across the transition
means that this quantity is not a suitable diagnostic of the Kondo-destruction QCP. 
The existence of the Kondo-destruction quantum phase transition can be seen through other properties, 
including the $\epsilon$-expansion with RG flow to the Kondo-destroyed state, the crossing and scaling
collapse of the Binder ratio, the divergence and scaling collapse of the fidelity susceptibility, 
a bifurcation in the NRG many-body spectra, and a vanishing crossover energy scale when the QCP is
approached from either side. These results provide evidence for a quantum phase transition at which
the amplitude of the static Kondo singlet goes continuously from nonzero in the Kondo-screened phase to zero
in the Kondo-destroyed phase. For example, the $\epsilon$-expansion RG shows that the Kondo-destroyed phase
has a vanishing Kondo coupling, signifying a vanishing amplitude for the static Kondo singlet.
However, along the trajectory {\it during the flow towards the fixed point}, the effective Kondo coupling
is nonzero and gives rise to the dynamical Kondo effect.

Third, the Kondo lattice model has been solved within the EDMFT in terms of an effective BFK model in the
presence of a local magnetic field \cite{si2001locally,si2003local,grempel2003locally,
jxzhu2003continuous,glossop2007magnetic,jxzhu2007zero}. Such studies have identified a continuous
Kondo-destruction quantum phase transition, where a self-consistently determined local magnetic field
rises continuously from zero as the system enters the Kondo-destroyed phase. The results presented here
imply that  $\langle {\bf S}_{f} \cdot {\bf s}_{c} \rangle$ will be smooth across the Kondo-destruction QCP 
in the Kondo lattice case. 

Fourth, we emphasize that in the Kondo lattice problem, the dynamical Kondo effect is crucial for
stabilizing the Kondo-destroyed phase, namely, the antiferromagnetic state with a small Fermi surface, 
referred to as AF$_{\text{S}}$ in the literature \cite{Si06.1}. If the dynamical Kondo effect is neglected,
AF$_{\text{S}}$ is not energetically favored. For instance, in variational quantum Monte Carlo studies
\cite{watanabe2007fermi}, the energy gain from the Kondo coupling only shows up in the form of a static
Kondo amplitude of the trial wave-function, which is zero in the AF$_{\text{S}}$ phase. The expectation value
$\langle {\bf S}_{f} \cdot {\bf s}_{c} \rangle$ vanishes in the AF$_{\text{S}}$ phase.
Thus, such a variational quantum Monte Carlo approach cannot capture the dynamical Kondo effect and 
the AF$_{\text{S}}$ is absent in this approach.

\section{Summary}
\label{sec:summary}

We have calculated the expectation value for the product of the local-moment and conduction-electron
spins in SU(2)-symmetric and Ising-anisotropic Bose-Fermi Anderson models where Kondo-destruction
quantum-critical points are unambiguously identified. Our key results are contained in Fig.\ \ref{fig:sfsc},
which shows this expectation value to vary continuously across the Kondo-destruction quantum critical
points. Through a spectral decomposition, this nonzero expectation value demonstrates the dynamical Kondo
effect that operates in the Kondo-destroyed phase. This dynamical Kondo effect is important for 
the stability of Kondo-destruction quantum criticality. In addition, it provides understanding of the 
enhanced effective mass in the Kondo-destroyed phase implicated in quantum-critical heavy-fermion materials
such as YbRh$_2$Si$_2$, Au-doped CeCu$_6$, CeRhIn$_5$, and Ce$_3$Pd$_{20}$Si$_6$.

\section{Acknowledgments}

We thank E.\ M.\ Nica for useful discussions.
This work was supported in part by the NSF under Grant Nos.\ DMR-{1920740}
(A.C., Q.S.) and
DMR-1508122 (K.I.), and by the Robert A. Welch Foundation C-1411 (H.H.). 
Computing time was supported in part by the Data Analysis and Visualization Cyberinfrastructure funded 
by NSF under grant OCI-0959097 and an IBM Shared University Research (SUR) Award at Rice University, 
and by the Extreme Science and Engineering
Discovery Environment (XSEDE) by NSF under Grant No.\ DMR-170109.
S.P.\ acknowledges funding from the Austrian Science Fund (project P29296-N27)
and from the European Union's Horizon 2020 research and innovation programme under grant agreement No.\ 824109.
Q.S.\ acknowledges hospitality and support under a Ulam Scholarship from the Center for Nonlinear Studies
at Los Alamos National Laboratory, and the hospitality of the Aspen Center for Physics, which is supported
by NSF Grant No.\ PHY-1607611.

\bibliography{dynamical_Kondo}

\begin{thebibliography}{50}
\expandafter\ifx\csname natexlab\endcsname\relax\def\natexlab#1{#1}\fi
\expandafter\ifx\csname bibnamefont\endcsname\relax
  \def\bibnamefont#1{#1}\fi
\expandafter\ifx\csname bibfnamefont\endcsname\relax
  \def\bibfnamefont#1{#1}\fi
\expandafter\ifx\csname citenamefont\endcsname\relax
  \def\citenamefont#1{#1}\fi
\expandafter\ifx\csname url\endcsname\relax
  \def\url#1{\texttt{#1}}\fi
\expandafter\ifx\csname urlprefix\endcsname\relax\def\urlprefix{URL }\fi
\providecommand{\bibinfo}[2]{#2}
\providecommand{\eprint}[2][]{\url{#2}}

\bibitem[{Spe(2013)}]{SpecialIssue2013}
\bibinfo{journal}{Special Issue on Quantum Criticality and Novel Phases, Phys.
  Stat. Sol. (b)} \textbf{\bibinfo{volume}{250}}, \bibinfo{pages}{417}
  (\bibinfo{year}{2013}).

\bibitem[{Spe(2010)}]{SpecialIssue2010}
\bibinfo{journal}{Special issue on Quantum Phase Transitions, J. Low Temp.
  Phys.} \textbf{\bibinfo{volume}{161}}, \bibinfo{pages}{1}
  (\bibinfo{year}{2010}).

\bibitem[{\citenamefont{Sachdev}(2011)}]{sachdev2011quantum}
\bibinfo{author}{\bibfnamefont{S.}~\bibnamefont{Sachdev}},
  \emph{\bibinfo{title}{Quantum Phase Transitions}}
  (\bibinfo{publisher}{Cambridge university press}, \bibinfo{year}{2011}).

\bibitem[{\citenamefont{Si and Steglich}(2010)}]{si2010heavy}
\bibinfo{author}{\bibfnamefont{Q.}~\bibnamefont{Si}} \bibnamefont{and}
  \bibinfo{author}{\bibfnamefont{F.}~\bibnamefont{Steglich}},
  \bibinfo{journal}{Science} \textbf{\bibinfo{volume}{329}},
  \bibinfo{pages}{1161} (\bibinfo{year}{2010}).

\bibitem[{\citenamefont{Coleman and Schofield}(2005)}]{Coleman-Nature}
\bibinfo{author}{\bibfnamefont{P.}~\bibnamefont{Coleman}} \bibnamefont{and}
  \bibinfo{author}{\bibfnamefont{A.~J.} \bibnamefont{Schofield}},
  \bibinfo{journal}{Nature} \textbf{\bibinfo{volume}{433}},
  \bibinfo{pages}{226} (\bibinfo{year}{2005}).

\bibitem[{\citenamefont{Stewart}(2001)}]{stewart2001non}
\bibinfo{author}{\bibfnamefont{G.~R.} \bibnamefont{Stewart}},
  \bibinfo{journal}{Rev. Mod. Phys.} \textbf{\bibinfo{volume}{73}},
  \bibinfo{pages}{797} (\bibinfo{year}{2001}).

\bibitem[{\citenamefont{Si and Paschen}(2013)}]{si2013quantum}
\bibinfo{author}{\bibfnamefont{Q.}~\bibnamefont{Si}} \bibnamefont{and}
  \bibinfo{author}{\bibfnamefont{S.}~\bibnamefont{Paschen}},
  \bibinfo{journal}{Phys. Stat Sol. (b)} \textbf{\bibinfo{volume}{250}},
  \bibinfo{pages}{425} (\bibinfo{year}{2013}).

\bibitem[{\citenamefont{Si et~al.}(2001)\citenamefont{Si, Rabello, Ingersent,
  and Smith}}]{si2001locally}
\bibinfo{author}{\bibfnamefont{Q.}~\bibnamefont{Si}},
  \bibinfo{author}{\bibfnamefont{S.}~\bibnamefont{Rabello}},
  \bibinfo{author}{\bibfnamefont{K.}~\bibnamefont{Ingersent}},
  \bibnamefont{and} \bibinfo{author}{\bibfnamefont{J.~L.} \bibnamefont{Smith}},
  \bibinfo{journal}{Nature} \textbf{\bibinfo{volume}{413}},
  \bibinfo{pages}{804} (\bibinfo{year}{2001}).

\bibitem[{\citenamefont{Coleman et~al.}(2001)\citenamefont{Coleman, P\'{e}pin,
  Si, and Ramazashvili}}]{Colemanetal}
\bibinfo{author}{\bibfnamefont{P.}~\bibnamefont{Coleman}},
  \bibinfo{author}{\bibfnamefont{C.}~\bibnamefont{P\'{e}pin}},
  \bibinfo{author}{\bibfnamefont{Q.}~\bibnamefont{Si}}, \bibnamefont{and}
  \bibinfo{author}{\bibfnamefont{R.}~\bibnamefont{Ramazashvili}},
  \bibinfo{journal}{J.~Phys.:~Cond.~Matt.} \textbf{\bibinfo{volume}{13}},
  \bibinfo{pages}{R723} (\bibinfo{year}{2001}).

\bibitem[{\citenamefont{Senthil et~al.}(2004)\citenamefont{Senthil, Vojta, and
  Sachdev}}]{senthil2004a}
\bibinfo{author}{\bibfnamefont{T.}~\bibnamefont{Senthil}},
  \bibinfo{author}{\bibfnamefont{M.}~\bibnamefont{Vojta}}, \bibnamefont{and}
  \bibinfo{author}{\bibfnamefont{S.}~\bibnamefont{Sachdev}},
  \bibinfo{journal}{Phys.~Rev.~B} \textbf{\bibinfo{volume}{69}},
  \bibinfo{pages}{035111} (\bibinfo{year}{2004}).

\bibitem[{\citenamefont{Hertz}(1976)}]{hertz1976quantum}
\bibinfo{author}{\bibfnamefont{J.~A.} \bibnamefont{Hertz}},
  \bibinfo{journal}{Phys. Rev. B} \textbf{\bibinfo{volume}{14}},
  \bibinfo{pages}{1165} (\bibinfo{year}{1976}).

\bibitem[{\citenamefont{Millis}(1993)}]{millis1993effect}
\bibinfo{author}{\bibfnamefont{A.~J.} \bibnamefont{Millis}},
  \bibinfo{journal}{Phys. Rev. B} \textbf{\bibinfo{volume}{48}},
  \bibinfo{pages}{7183} (\bibinfo{year}{1993}).

\bibitem[{\citenamefont{Moriya}(2012)}]{moriya2012spin}
\bibinfo{author}{\bibfnamefont{T.}~\bibnamefont{Moriya}},
  \emph{\bibinfo{title}{Spin fluctuations in itinerant electron magnetism}},
  vol.~\bibinfo{volume}{56} (\bibinfo{publisher}{Springer Science \& Business
  Media}, \bibinfo{year}{2012}).

\bibitem[{\citenamefont{Schr{\"o}der et~al.}(2000)\citenamefont{Schr{\"o}der,
  Aeppli, Coldea, Adams, Stockert, L{\"o}hneysen, Bucher, Ramazashvili, and
  Coleman}}]{schroder2000onset}
\bibinfo{author}{\bibfnamefont{A.}~\bibnamefont{Schr{\"o}der}},
  \bibinfo{author}{\bibfnamefont{G.}~\bibnamefont{Aeppli}},
  \bibinfo{author}{\bibfnamefont{R.}~\bibnamefont{Coldea}},
  \bibinfo{author}{\bibfnamefont{M.}~\bibnamefont{Adams}},
  \bibinfo{author}{\bibfnamefont{O.}~\bibnamefont{Stockert}},
  \bibinfo{author}{\bibfnamefont{H.~v.} \bibnamefont{L{\"o}hneysen}},
  \bibinfo{author}{\bibfnamefont{E.}~\bibnamefont{Bucher}},
  \bibinfo{author}{\bibfnamefont{R.}~\bibnamefont{Ramazashvili}},
  \bibnamefont{and} \bibinfo{author}{\bibfnamefont{P.}~\bibnamefont{Coleman}},
  \bibinfo{journal}{Nature} \textbf{\bibinfo{volume}{407}},
  \bibinfo{pages}{351} (\bibinfo{year}{2000}).

\bibitem[{\citenamefont{Prochaska et~al.}(2018)\citenamefont{Prochaska, Li,
  MacFarland, Andrews, Bonta, Bianco, Yazdi, Schrenk, Detz, Limbeck
  et~al.}}]{prochaska2018singular}
\bibinfo{author}{\bibfnamefont{L.}~\bibnamefont{Prochaska}},
  \bibinfo{author}{\bibfnamefont{X.}~\bibnamefont{Li}},
  \bibinfo{author}{\bibfnamefont{D.~C.} \bibnamefont{MacFarland}},
  \bibinfo{author}{\bibfnamefont{A.~M.} \bibnamefont{Andrews}},
  \bibinfo{author}{\bibfnamefont{M.}~\bibnamefont{Bonta}},
  \bibinfo{author}{\bibfnamefont{E.~F.} \bibnamefont{Bianco}},
  \bibinfo{author}{\bibfnamefont{S.}~\bibnamefont{Yazdi}},
  \bibinfo{author}{\bibfnamefont{W.}~\bibnamefont{Schrenk}},
  \bibinfo{author}{\bibfnamefont{H.}~\bibnamefont{Detz}},
  \bibinfo{author}{\bibfnamefont{A.}~\bibnamefont{Limbeck}},
  \bibnamefont{et~al.}, \bibinfo{journal}{arXiv preprint arXiv:1808.02296}
  (\bibinfo{year}{2018}).

\bibitem[{\citenamefont{Paschen et~al.}(2004)\citenamefont{Paschen,
  L{\"u}hmann, Wirth, Gegenwart, Trovarelli, Geibel, Steglich, Coleman, and
  Si}}]{paschen2004hall}
\bibinfo{author}{\bibfnamefont{S.}~\bibnamefont{Paschen}},
  \bibinfo{author}{\bibfnamefont{T.}~\bibnamefont{L{\"u}hmann}},
  \bibinfo{author}{\bibfnamefont{S.}~\bibnamefont{Wirth}},
  \bibinfo{author}{\bibfnamefont{P.}~\bibnamefont{Gegenwart}},
  \bibinfo{author}{\bibfnamefont{O.}~\bibnamefont{Trovarelli}},
  \bibinfo{author}{\bibfnamefont{C.}~\bibnamefont{Geibel}},
  \bibinfo{author}{\bibfnamefont{F.}~\bibnamefont{Steglich}},
  \bibinfo{author}{\bibfnamefont{P.}~\bibnamefont{Coleman}}, \bibnamefont{and}
  \bibinfo{author}{\bibfnamefont{Q.}~\bibnamefont{Si}},
  \bibinfo{journal}{Nature} \textbf{\bibinfo{volume}{432}},
  \bibinfo{pages}{881} (\bibinfo{year}{2004}).

\bibitem[{\citenamefont{Friedemann et~al.}(2010)\citenamefont{Friedemann,
  Oeschler, Wirth, Krellner, Geibel, Steglich, Paschen, Kirchner, and
  Si}}]{Fri10.2}
\bibinfo{author}{\bibfnamefont{S.}~\bibnamefont{Friedemann}},
  \bibinfo{author}{\bibfnamefont{N.}~\bibnamefont{Oeschler}},
  \bibinfo{author}{\bibfnamefont{S.}~\bibnamefont{Wirth}},
  \bibinfo{author}{\bibfnamefont{C.}~\bibnamefont{Krellner}},
  \bibinfo{author}{\bibfnamefont{C.}~\bibnamefont{Geibel}},
  \bibinfo{author}{\bibfnamefont{F.}~\bibnamefont{Steglich}},
  \bibinfo{author}{\bibfnamefont{S.}~\bibnamefont{Paschen}},
  \bibinfo{author}{\bibfnamefont{S.}~\bibnamefont{Kirchner}}, \bibnamefont{and}
  \bibinfo{author}{\bibfnamefont{Q.}~\bibnamefont{Si}},
  \bibinfo{journal}{{Proc.\ Natl.\ Acad.\ Sci.\ USA}}
  \textbf{\bibinfo{volume}{{107}}}, \bibinfo{pages}{14547}
  (\bibinfo{year}{2010}).

\bibitem[{\citenamefont{Shishido et~al.}(2005)\citenamefont{Shishido, Settai,
  Harima, and {\=O}nuki}}]{shishido2005drastic}
\bibinfo{author}{\bibfnamefont{H.}~\bibnamefont{Shishido}},
  \bibinfo{author}{\bibfnamefont{R.}~\bibnamefont{Settai}},
  \bibinfo{author}{\bibfnamefont{H.}~\bibnamefont{Harima}}, \bibnamefont{and}
  \bibinfo{author}{\bibfnamefont{Y.}~\bibnamefont{{\=O}nuki}},
  \bibinfo{journal}{J. Phys. Soc. Japan} \textbf{\bibinfo{volume}{74}},
  \bibinfo{pages}{1103} (\bibinfo{year}{2005}).

\bibitem[{\citenamefont{Custers et~al.}(2012)\citenamefont{Custers, Lorenzer,
  M{\"u}ller, Prokofiev, Sidorenko, Winkler, Strydom, Shimura, Sakakibara, Yu
  et~al.}}]{custers2012destruction}
\bibinfo{author}{\bibfnamefont{J.}~\bibnamefont{Custers}},
  \bibinfo{author}{\bibfnamefont{K.~A.} \bibnamefont{Lorenzer}},
  \bibinfo{author}{\bibfnamefont{M.}~\bibnamefont{M{\"u}ller}},
  \bibinfo{author}{\bibfnamefont{A.}~\bibnamefont{Prokofiev}},
  \bibinfo{author}{\bibfnamefont{A.}~\bibnamefont{Sidorenko}},
  \bibinfo{author}{\bibfnamefont{H.}~\bibnamefont{Winkler}},
  \bibinfo{author}{\bibfnamefont{A.~M.} \bibnamefont{Strydom}},
  \bibinfo{author}{\bibfnamefont{Y.}~\bibnamefont{Shimura}},
  \bibinfo{author}{\bibfnamefont{T.}~\bibnamefont{Sakakibara}},
  \bibinfo{author}{\bibfnamefont{R.}~\bibnamefont{Yu}}, \bibnamefont{et~al.},
  \bibinfo{journal}{Nat. Mater.} \textbf{\bibinfo{volume}{11}},
  \bibinfo{pages}{189} (\bibinfo{year}{2012}).

\bibitem[{\citenamefont{Martelli et~al.}(2017)\citenamefont{Martelli, Cai,
  Nica, Taupin, Prokofiev, Liu, Lai, Yu, K{\"u}chler, Strydom
  et~al.}}]{martelli2017sequential}
\bibinfo{author}{\bibfnamefont{V.}~\bibnamefont{Martelli}},
  \bibinfo{author}{\bibfnamefont{A.}~\bibnamefont{Cai}},
  \bibinfo{author}{\bibfnamefont{E.~M.} \bibnamefont{Nica}},
  \bibinfo{author}{\bibfnamefont{M.}~\bibnamefont{Taupin}},
  \bibinfo{author}{\bibfnamefont{A.}~\bibnamefont{Prokofiev}},
  \bibinfo{author}{\bibfnamefont{C.-C.} \bibnamefont{Liu}},
  \bibinfo{author}{\bibfnamefont{H.-H.} \bibnamefont{Lai}},
  \bibinfo{author}{\bibfnamefont{R.}~\bibnamefont{Yu}},
  \bibinfo{author}{\bibfnamefont{R.}~\bibnamefont{K{\"u}chler}},
  \bibinfo{author}{\bibfnamefont{A.~M.} \bibnamefont{Strydom}},
  \bibnamefont{et~al.}, \bibinfo{journal}{arXiv preprint arXiv:1709.09376}
  (\bibinfo{year}{2017}).

\bibitem[{\citenamefont{Si et~al.}(2003)\citenamefont{Si, Rabello, Ingersent,
  and Smith}}]{si2003local}
\bibinfo{author}{\bibfnamefont{Q.}~\bibnamefont{Si}},
  \bibinfo{author}{\bibfnamefont{S.}~\bibnamefont{Rabello}},
  \bibinfo{author}{\bibfnamefont{K.}~\bibnamefont{Ingersent}},
  \bibnamefont{and} \bibinfo{author}{\bibfnamefont{J.~L.} \bibnamefont{Smith}},
  \bibinfo{journal}{Phys. Rev. B} \textbf{\bibinfo{volume}{68}},
  \bibinfo{pages}{115103} (\bibinfo{year}{2003}).

\bibitem[{\citenamefont{Grempel and Si}(2003)}]{grempel2003locally}
\bibinfo{author}{\bibfnamefont{D.~R.} \bibnamefont{Grempel}} \bibnamefont{and}
  \bibinfo{author}{\bibfnamefont{Q.}~\bibnamefont{Si}}, \bibinfo{journal}{Phys.
  Rev. Lett.} \textbf{\bibinfo{volume}{91}}, \bibinfo{pages}{026401}
  (\bibinfo{year}{2003}).

\bibitem[{\citenamefont{Zhu et~al.}(2003{\natexlab{a}})\citenamefont{Zhu,
  Grempel, and Si}}]{jxzhu2003continuous}
\bibinfo{author}{\bibfnamefont{J.-X.} \bibnamefont{Zhu}},
  \bibinfo{author}{\bibfnamefont{D.~R.} \bibnamefont{Grempel}},
  \bibnamefont{and} \bibinfo{author}{\bibfnamefont{Q.}~\bibnamefont{Si}},
  \bibinfo{journal}{Phys. Rev. Lett.} \textbf{\bibinfo{volume}{91}},
  \bibinfo{pages}{156404} (\bibinfo{year}{2003}{\natexlab{a}}).

\bibitem[{\citenamefont{Glossop and
  Ingersent}(2007{\natexlab{a}})}]{glossop2007magnetic}
\bibinfo{author}{\bibfnamefont{M.~T.} \bibnamefont{Glossop}} \bibnamefont{and}
  \bibinfo{author}{\bibfnamefont{K.}~\bibnamefont{Ingersent}},
  \bibinfo{journal}{Phys. Rev. Lett.} \textbf{\bibinfo{volume}{99}},
  \bibinfo{pages}{227203} (\bibinfo{year}{2007}{\natexlab{a}}).

\bibitem[{\citenamefont{Zhu et~al.}(2007)\citenamefont{Zhu, Kirchner, Bulla,
  and Si}}]{jxzhu2007zero}
\bibinfo{author}{\bibfnamefont{J.-X.} \bibnamefont{Zhu}},
  \bibinfo{author}{\bibfnamefont{S.}~\bibnamefont{Kirchner}},
  \bibinfo{author}{\bibfnamefont{R.}~\bibnamefont{Bulla}}, \bibnamefont{and}
  \bibinfo{author}{\bibfnamefont{Q.}~\bibnamefont{Si}}, \bibinfo{journal}{Phys.
  Rev. Lett.} \textbf{\bibinfo{volume}{99}}, \bibinfo{pages}{227204}
  (\bibinfo{year}{2007}).

\bibitem[{\citenamefont{Si et~al.}(2014)\citenamefont{Si, Pixley, Nica,
  Yamamoto, Goswami, Yu, and Kirchner}}]{si2014kondo}
\bibinfo{author}{\bibfnamefont{Q.}~\bibnamefont{Si}},
  \bibinfo{author}{\bibfnamefont{J.~H.} \bibnamefont{Pixley}},
  \bibinfo{author}{\bibfnamefont{E.}~\bibnamefont{Nica}},
  \bibinfo{author}{\bibfnamefont{S.~J.} \bibnamefont{Yamamoto}},
  \bibinfo{author}{\bibfnamefont{P.}~\bibnamefont{Goswami}},
  \bibinfo{author}{\bibfnamefont{R.}~\bibnamefont{Yu}}, \bibnamefont{and}
  \bibinfo{author}{\bibfnamefont{S.}~\bibnamefont{Kirchner}},
  \bibinfo{journal}{J. Phys. Soc. Japan} \textbf{\bibinfo{volume}{83}},
  \bibinfo{pages}{061005} (\bibinfo{year}{2014}).

\bibitem[{\citenamefont{Zhu et~al.}(2003{\natexlab{b}})\citenamefont{Zhu,
  Grempel, and Si}}]{zhu2003continuous}
\bibinfo{author}{\bibfnamefont{J.-X.} \bibnamefont{Zhu}},
  \bibinfo{author}{\bibfnamefont{D.}~\bibnamefont{Grempel}}, \bibnamefont{and}
  \bibinfo{author}{\bibfnamefont{Q.}~\bibnamefont{Si}}, \bibinfo{journal}{Phys.
  Rev. Lett.} \textbf{\bibinfo{volume}{91}}, \bibinfo{pages}{156404}
  (\bibinfo{year}{2003}{\natexlab{b}}).

\bibitem[{\citenamefont{Gegenwart et~al.}(2002)\citenamefont{Gegenwart,
  Custers, Geibel, Neumaier, Tayama, Tenya, Trovarelli, and
  Steglich}}]{Geg02.1}
\bibinfo{author}{\bibfnamefont{P.}~\bibnamefont{Gegenwart}},
  \bibinfo{author}{\bibfnamefont{J.}~\bibnamefont{Custers}},
  \bibinfo{author}{\bibfnamefont{C.}~\bibnamefont{Geibel}},
  \bibinfo{author}{\bibfnamefont{K.}~\bibnamefont{Neumaier}},
  \bibinfo{author}{\bibfnamefont{T.}~\bibnamefont{Tayama}},
  \bibinfo{author}{\bibfnamefont{K.}~\bibnamefont{Tenya}},
  \bibinfo{author}{\bibfnamefont{O.}~\bibnamefont{Trovarelli}},
  \bibnamefont{and} \bibinfo{author}{\bibfnamefont{F.}~\bibnamefont{Steglich}},
  \bibinfo{journal}{{Phys.\ Rev.\ Lett.}} \textbf{\bibinfo{volume}{89}},
  \bibinfo{pages}{056402} (\bibinfo{year}{2002}).

\bibitem[{\citenamefont{{v.\ L\"ohneysen} et~al.}(1994)\citenamefont{{v.\
  L\"ohneysen}, Pietrus, Portisch, Schlager, Schr{\"o}der, Sieck, and
  Trappmann}}]{Loe94.1}
\bibinfo{author}{\bibfnamefont{H.}~\bibnamefont{{v.\ L\"ohneysen}}},
  \bibinfo{author}{\bibfnamefont{T.}~\bibnamefont{Pietrus}},
  \bibinfo{author}{\bibfnamefont{G.}~\bibnamefont{Portisch}},
  \bibinfo{author}{\bibfnamefont{H.~G.} \bibnamefont{Schlager}},
  \bibinfo{author}{\bibfnamefont{A.}~\bibnamefont{Schr{\"o}der}},
  \bibinfo{author}{\bibfnamefont{M.}~\bibnamefont{Sieck}}, \bibnamefont{and}
  \bibinfo{author}{\bibfnamefont{T.}~\bibnamefont{Trappmann}},
  \bibinfo{journal}{{Phys.\ Rev.\ Lett.}} \textbf{\bibinfo{volume}{72}},
  \bibinfo{pages}{3262} (\bibinfo{year}{1994}).

\bibitem[{\citenamefont{Hewson}(1997)}]{Hew97.1}
\bibinfo{author}{\bibfnamefont{A.~C.} \bibnamefont{Hewson}},
  \emph{\bibinfo{title}{{The Kondo Problem to Heavy Fermions}}}
  (\bibinfo{publisher}{Cambridge University Press},
  \bibinfo{address}{Cambridge}, \bibinfo{year}{1997}).

\bibitem[{\citenamefont{Cai and Si}(2019)}]{cai2019}
\bibinfo{author}{\bibfnamefont{A.}~\bibnamefont{Cai}} \bibnamefont{and}
  \bibinfo{author}{\bibfnamefont{Q.}~\bibnamefont{Si}}, \bibinfo{journal}{arXiv
  preprint arXiv:1902.10094}  (\bibinfo{year}{2019}).

\bibitem[{\citenamefont{Otsuki}(2013)}]{otsuki2013spin}
\bibinfo{author}{\bibfnamefont{J.}~\bibnamefont{Otsuki}},
  \bibinfo{journal}{Phys. Rev. B} \textbf{\bibinfo{volume}{87}},
  \bibinfo{pages}{125102} (\bibinfo{year}{2013}).

\bibitem[{\citenamefont{Si and Smith}(1996)}]{si1996kosterlitz}
\bibinfo{author}{\bibfnamefont{Q.}~\bibnamefont{Si}} \bibnamefont{and}
  \bibinfo{author}{\bibfnamefont{J.~L.} \bibnamefont{Smith}},
  \bibinfo{journal}{Phys. Rev. Lett.} \textbf{\bibinfo{volume}{77}},
  \bibinfo{pages}{3391} (\bibinfo{year}{1996}).

\bibitem[{\citenamefont{Smith and Si}(2000)}]{smith2000spatial}
\bibinfo{author}{\bibfnamefont{J.~L.} \bibnamefont{Smith}} \bibnamefont{and}
  \bibinfo{author}{\bibfnamefont{Q.}~\bibnamefont{Si}}, \bibinfo{journal}{Phys.
  Rev. B} \textbf{\bibinfo{volume}{61}}, \bibinfo{pages}{5184}
  (\bibinfo{year}{2000}).

\bibitem[{\citenamefont{Chitra and Kotliar}(2000)}]{chitra2000effect}
\bibinfo{author}{\bibfnamefont{R.}~\bibnamefont{Chitra}} \bibnamefont{and}
  \bibinfo{author}{\bibfnamefont{G.}~\bibnamefont{Kotliar}},
  \bibinfo{journal}{Phys. Rev. Lett.} \textbf{\bibinfo{volume}{84}},
  \bibinfo{pages}{3678} (\bibinfo{year}{2000}).

\bibitem[{\citenamefont{Zhu and Si}(2002)}]{zhu2002}
\bibinfo{author}{\bibfnamefont{L.}~\bibnamefont{Zhu}} \bibnamefont{and}
  \bibinfo{author}{\bibfnamefont{Q.}~\bibnamefont{Si}}, \bibinfo{journal}{Phys.
  Rev. B} \textbf{\bibinfo{volume}{66}}, \bibinfo{pages}{024426}
  (\bibinfo{year}{2002}).

\bibitem[{\citenamefont{Zar{\'a}nd and Demler}(2002)}]{zarand2002quantum}
\bibinfo{author}{\bibfnamefont{G.}~\bibnamefont{Zar{\'a}nd}} \bibnamefont{and}
  \bibinfo{author}{\bibfnamefont{E.}~\bibnamefont{Demler}},
  \bibinfo{journal}{Phys. Rev. B} \textbf{\bibinfo{volume}{66}},
  \bibinfo{pages}{024427} (\bibinfo{year}{2002}).

\bibitem[{\citenamefont{Smith and Si}(1999)}]{smith1999non}
\bibinfo{author}{\bibfnamefont{J.~L.} \bibnamefont{Smith}} \bibnamefont{and}
  \bibinfo{author}{\bibfnamefont{Q.}~\bibnamefont{Si}},
  \bibinfo{journal}{Europhys. Lett.} \textbf{\bibinfo{volume}{45}},
  \bibinfo{pages}{228} (\bibinfo{year}{1999}).

\bibitem[{\citenamefont{Sengupta}(2000)}]{sengupta2000spin}
\bibinfo{author}{\bibfnamefont{A.~M.} \bibnamefont{Sengupta}},
  \bibinfo{journal}{Phys. Rev. B} \textbf{\bibinfo{volume}{61}},
  \bibinfo{pages}{4041} (\bibinfo{year}{2000}).

\bibitem[{\citenamefont{You et~al.}(2007)\citenamefont{You, Li, and
  Gu}}]{you2007fidelity}
\bibinfo{author}{\bibfnamefont{W.-L.} \bibnamefont{You}},
  \bibinfo{author}{\bibfnamefont{Y.-W.} \bibnamefont{Li}}, \bibnamefont{and}
  \bibinfo{author}{\bibfnamefont{S.-J.} \bibnamefont{Gu}},
  \bibinfo{journal}{Phys. Rev. E} \textbf{\bibinfo{volume}{76}},
  \bibinfo{pages}{022101} (\bibinfo{year}{2007}).

\bibitem[{\citenamefont{Albuquerque et~al.}(2010)\citenamefont{Albuquerque,
  Alet, Sire, and Capponi}}]{albuquerque2010}
\bibinfo{author}{\bibfnamefont{A.~F.} \bibnamefont{Albuquerque}},
  \bibinfo{author}{\bibfnamefont{F.}~\bibnamefont{Alet}},
  \bibinfo{author}{\bibfnamefont{C.}~\bibnamefont{Sire}}, \bibnamefont{and}
  \bibinfo{author}{\bibfnamefont{S.}~\bibnamefont{Capponi}},
  \bibinfo{journal}{Phys. Rev. B} \textbf{\bibinfo{volume}{81}},
  \bibinfo{pages}{064418} (\bibinfo{year}{2010}).

\bibitem[{\citenamefont{Wang et~al.}(2015)\citenamefont{Wang, Liu,
  Imri{\v{s}}ka, Ma, and Troyer}}]{wang2015fidelity}
\bibinfo{author}{\bibfnamefont{L.}~\bibnamefont{Wang}},
  \bibinfo{author}{\bibfnamefont{Y.-H.} \bibnamefont{Liu}},
  \bibinfo{author}{\bibfnamefont{J.}~\bibnamefont{Imri{\v{s}}ka}},
  \bibinfo{author}{\bibfnamefont{P.~N.} \bibnamefont{Ma}}, \bibnamefont{and}
  \bibinfo{author}{\bibfnamefont{M.}~\bibnamefont{Troyer}},
  \bibinfo{journal}{Phys. Rev. X} \textbf{\bibinfo{volume}{5}},
  \bibinfo{pages}{031007} (\bibinfo{year}{2015}).

\bibitem[{\citenamefont{Beach et~al.}(2005)\citenamefont{Beach, Wang, and
  Sandvik}}]{beach2005data}
\bibinfo{author}{\bibfnamefont{K.~S.~D.} \bibnamefont{Beach}},
  \bibinfo{author}{\bibfnamefont{L.}~\bibnamefont{Wang}}, \bibnamefont{and}
  \bibinfo{author}{\bibfnamefont{A.~W.} \bibnamefont{Sandvik}},
  \bibinfo{journal}{arXiv preprint cond-mat/0505194}  (\bibinfo{year}{2005}).

\bibitem[{\citenamefont{Houdayer and Hartmann}(2004)}]{houdayer2004low}
\bibinfo{author}{\bibfnamefont{J.}~\bibnamefont{Houdayer}} \bibnamefont{and}
  \bibinfo{author}{\bibfnamefont{A.~K.} \bibnamefont{Hartmann}},
  \bibinfo{journal}{Phys. Rev. B} \textbf{\bibinfo{volume}{70}},
  \bibinfo{pages}{014418} (\bibinfo{year}{2004}).

\bibitem[{\citenamefont{Pixley et~al.}(2013)\citenamefont{Pixley, Kirchner,
  Ingersent, and Si}}]{pixley2013quantum}
\bibinfo{author}{\bibfnamefont{J.~H.} \bibnamefont{Pixley}},
  \bibinfo{author}{\bibfnamefont{S.}~\bibnamefont{Kirchner}},
  \bibinfo{author}{\bibfnamefont{K.}~\bibnamefont{Ingersent}},
  \bibnamefont{and} \bibinfo{author}{\bibfnamefont{Q.}~\bibnamefont{Si}},
  \bibinfo{journal}{Phys. Rev. B} \textbf{\bibinfo{volume}{88}},
  \bibinfo{pages}{245111} (\bibinfo{year}{2013}).

\bibitem[{\citenamefont{Binder}(1981)}]{binder1981finite}
\bibinfo{author}{\bibfnamefont{K.}~\bibnamefont{Binder}}, \bibinfo{journal}{Z.
  Phys. B} \textbf{\bibinfo{volume}{43}}, \bibinfo{pages}{119}
  (\bibinfo{year}{1981}).

\bibitem[{\citenamefont{Glossop and Ingersent}(2005)}]{glossop2005bfk}
\bibinfo{author}{\bibfnamefont{M.~T.} \bibnamefont{Glossop}} \bibnamefont{and}
  \bibinfo{author}{\bibfnamefont{K.}~\bibnamefont{Ingersent}},
  \bibinfo{journal}{Phys. Rev. Lett.} \textbf{\bibinfo{volume}{95}},
  \bibinfo{pages}{067202} (\bibinfo{year}{2005}).

\bibitem[{\citenamefont{Glossop and
  Ingersent}(2007{\natexlab{b}})}]{glossop2007bfk}
\bibinfo{author}{\bibfnamefont{M.~T.} \bibnamefont{Glossop}} \bibnamefont{and}
  \bibinfo{author}{\bibfnamefont{K.}~\bibnamefont{Ingersent}},
  \bibinfo{journal}{Phys. Rev. B} \textbf{\bibinfo{volume}{75}},
  \bibinfo{pages}{104410} (\bibinfo{year}{2007}{\natexlab{b}}).

\bibitem[{\citenamefont{Si}(2006)}]{Si06.1}
\bibinfo{author}{\bibfnamefont{Q.}~\bibnamefont{Si}}, \bibinfo{journal}{Physica
  B} \textbf{\bibinfo{volume}{378}}, \bibinfo{pages}{23}
  (\bibinfo{year}{2006}).

\bibitem[{\citenamefont{Watanabe and Ogata}(2007)}]{watanabe2007fermi}
\bibinfo{author}{\bibfnamefont{H.}~\bibnamefont{Watanabe}} \bibnamefont{and}
  \bibinfo{author}{\bibfnamefont{M.}~\bibnamefont{Ogata}},
  \bibinfo{journal}{Phys. Rev. Lett.} \textbf{\bibinfo{volume}{99}},
  \bibinfo{pages}{136401} (\bibinfo{year}{2007}).

\end{thebibliography}

\end{document}